\documentclass[onecolumn,aps,prd,superscriptaddress,floats,floatfix,showkeys,notitlepage]{revtex4}

\usepackage{graphicx,amssymb,amsmath,epsfig,mathtools,caption,csquotes} 
\usepackage{palatino}
\usepackage{xcolor}
\usepackage{hyperref}
\hypersetup{colorlinks=true,linkcolor=blue,urlcolor=green,citecolor=red}
\usepackage{tabularx}
\usepackage{booktabs}
\usepackage{multirow}
\usepackage{array}
\begin{document}

\title{Thermodynamic Topological Classifications of Well-Known Black Holes}

\author{Aqsa Mehmood}
\email{aqsamehmood0898@gmail.com},
\affiliation{Department of Mathematics, Faculty of Science, University of Okara, Okara 56130, Pakistan.}

\author{M. Umair Shahzad}
\email{mushahzad@uo.edu.pk;m.u.shahzad@ucp.edu.pk}
\affiliation{Department of Mathematics, Faculty of Science, University of Okara, Okara 56130, Pakistan.}

\date{\today}

\begin{abstract}
In this work, we investigate the thermodynamic properties of black holes (BHs) that have non-trivial topological features in their phase diagrams. We consider three different models of BHs: (1) a class of BHs in dRGT massive gravity, which adds a mass term to general relativity; (2) a class of BHs in 5D Yang-Mills massive gravity, which combines dRGT massive gravity with a non-Abelian gauge field; and (3) a D-dimensional RN-AdS BH surrounded by Quintessence and a cloud of strings, which are strange forms of matter that change the thermodynamics of the BH. Our goal is to find the critical points of these BHs, which provide the location of first-order phase transitions, and figure out their corresponding topological charges. Topological charges are numbers that show how complicated the BH topology is. Then, we look at these BHs as topological defects in the thermodynamic domain, which is the space of thermodynamic variables like pressure and temperature. We calculate winding numbers to analyse topology on a global and local scale at these defects, which are integers that indicate how many times a curve encircling the defect wraps around the origin. Our analysis reveals that the total topological charge is either equal to 0 or 1 for all models, meaning that the BHs have either a trivial or simple topology. In some cases, we see that the BHs' topology belongs to a different thermodynamic topological class. This means that the BHs can go through topological phase transitions.
\end{abstract}

\maketitle
\section{Introduction}
First, the light rings are th regions where photons are in unstable circular orbits around the BH have been studied in \cite{a1,a2,a3,a4}. The existence of these light rings could be more proof that BHs exist by creating a series of concentric rings in the shadow image of the BH. Recently, this active area of research has also expanded  the study of timelike circular orbits, which are the stable paths of heavy particles around the BH \cite{a5,a6}. These orbits
could reveal the properties of the BH’s spacetime, such as its mass, spin, and charge.\\

The second aspect of our study is the classification of a wide range of BHs based on their thermodynamic and topological characteristics \cite{a7}-\cite{a26}, which is a novel approach to classifying BHs based on their thermodynamic behaviour. This method uses topological tools, like winding numbers and topological charges, to describe the phase diagrams of BHs. These diagrams show how thermodynamic quantities, like entropy and temperature, change in relation to thermodynamic variables, like pressure and volume. Using this method, we can find the critical points in BHs where phase changes between different BH phases happen and figure out what their topological nature is. We can also distinguish different classes of BHs that have different topological features in their phase diagrams. This approach can reveal new insights into the thermodynamics and geometry of BHs. The topological method proposed in \cite{a2} has been successfully used to examine the topological properties of various renowned BH solutions \cite{a14}-\cite{a27}.\\

We follow the procedure of Refs. \cite{a8} and assign topological charges to the critical points of BHs. The topological charge is a quantity that measures the degree of non-triviality of the BH topology. It can be calculated via the following steps:
 \begin{itemize}
     \item First, we find the critical temperature of a BH, which is the temperature at which the BH undergoes a phase transition between different phases. The critical temperature is given by the condition
     \begin{equation}\label{1}
   ( \frac{\partial T}{\partial S} )_{P....} = 0.
\end{equation}
\item Second, we define a new “thermodynamic potential” that can be identified as the critical points. The thermodynamic potential is given by
\begin{equation}\label{2}
    \Phi = \frac{1}{sin\theta}\tilde{T}(S,...)
\end{equation}
where $\tilde T$ is the BH temperature \cite{a7}. The factor of $1/\sin \theta$ is an auxiliary factor that simplifies the topology of the critical points \cite{a8}.
\item Third, we define a vector field $\phi^a = (\phi^S,\phi^\theta)$ that represents the gradient of the thermodynamic potential regarding entropy and an angular variable $\theta$. The vector field is given by 
\begin{equation}\label{3}
    \phi^S = (\frac{\partial \Phi}{\partial S})_{\theta....}, \phi^\theta = (\frac{\partial \Phi}{\partial \theta})_{S....}
\end{equation}
\item Fourth, we compute the topological current $j^{\mu}$, which is a quantity that captures the flow of the vector field around the critical points. The topological current is given by
\begin{equation}\label{4} j^\mu = \frac{1}{2\pi}\epsilon^{\mu\nu\lambda}\epsilon_{ab}\partial_\nu n^a \partial_ \lambda n^b. \end{equation}
\item Next, we introduce a vector field $n^a = \phi^a/||\phi||$, where $\phi^a$ are the components of the gradient of a thermodynamic potential $\phi$ that depends on the entropy $S$ and an angular variable $\theta$. We also extend the (S,$\theta$) space to $x^\mu = (t, S,\theta)$, where $t$ is time-like variable. We find the critical points of the system, which are the points where the vector field vanishes. These points correspond to the phase transitions between different BH phases. We also verify that the vector field satisfies the conservation equation $\partial_\mu j^\mu = 0$.
\item Finally, we construct the topological charge Q by integrating the topological current over a closed surface $\sum$ that encloses a critical point. The topological charge is given by 
\begin{equation}\label{5}
    Q = \frac{1}{2\pi}\int_\Sigma j^\mu d^2 \Sigma_\mu = \sum_i w_i,
\end{equation} where $d^2\Sigma_\mu = \sigma_\mu d^2\Sigma$ is the surface element with $\sigma_\mu$  being the unit normal to $\Sigma$, and $w_i$ is the winding number of the $ith$ critical point. The winding number is an integer that indicates how many times the vector field wraps around the origin. The conventional critical point has $Q=-1$, while the unstable novel point has $Q=1$ \cite{a8}, \cite{a21}.
\end{itemize}
We apply this method to various classes of BHs and classify them according to their topological charges. We also compare our results with other approaches that use topology to study BH thermodynamics. Next, the generalized free energy $F$ is given by \begin{equation}\label{6}
    F = E-S/\tau
\end{equation}
where E is the energy, \& S is the entropy and $\tau$ is a time-like quantity \cite{a21}. Then, define a vector field $\phi$ given by
\begin{equation}\label{7}
\phi = (\frac{\partial F}{\partial r_+},-\cot\Theta \csc\Theta)
\end{equation}
where $r_+$ is the horizon radius and $\Theta$ is an angular variable.
The zero points of the vector $\phi$ are determined when $\Theta$ = $\pi/2$. The unit vector defined as follows:
\begin{equation}\label{8}
    n^\gamma = \frac{\phi^\gamma}{\|\phi\|}    (\gamma = 1,2)
\end{equation}
 {$ \phi^1  = \phi^{r_+},   \phi^2  = \phi^\Theta$}\\
 The zero points of this vector field correspond to specific values of $\tau$ and have their own winding numbers. By adding up these winding numbers for each BH branch, they obtain the topological number of a BH which is used in various BH solutions \cite{a12}-\cite{a26}.\\

In this research, our goal is to study the topological properties of BHs that have non-trivial features in their phase diagrams. We use the following steps to achieve this goal:
\begin{itemize}
    \item  First, we find the critical points of BHs, which are the points where the BHs undergo phase transitions between different phases. We also calculate their topological charges, which are quantities that measure the degree of non-triviality of the BH topology.
    \item Second, we classify the critical points into two types: conventional and novel. The conventional critical points have negative topological charges and indicate the presence of first-order phase transitions near them. The novel critical points have positive topological charges and do not imply any phase transition near them.
    \item Third, we treat the BHs as topological defects in the thermodynamic space, which is the space of thermodynamic variables such as temperature and pressure. We determine their topological numbers, which are sums of winding numbers of all defects. The winding number is an integer that indicates how many times a curve encircling a defect wraps around the origin. We also identify the specific locations in the thermodynamic space where the BHs are created and destroyed.
\end{itemize}

\section{A class of black hole in dRGT massive gravity}
 In the framework of dRGT massive gravity, a BH solution with an electric charge \cite{r1} is defined as
 \begin{equation}\label{9}
    f(r_+) = 1-\,{\frac {2M}{r_+}}+\frac{\lambda}{3}{r_+}^{2}+\eta\,r_++\zeta +{\frac {{Q}^{
2}}{{r_+}^{2}}}
 \end{equation}
 where
 \begin{equation}\label{10}
     \lambda = 3{m_g}^{2} \left( 1+\,\alpha+\,\beta \right)
 \end{equation}
 \begin{equation}\label{11}
     \eta=-c{m_g}^{2} \left( 1+2\,\alpha+3\,\beta \right) 
 \end{equation}
 \begin{equation}\label{12}
     \zeta ={c}^{2}{m_g}^{2} \left( \alpha+3\,\beta \right) 
 \end{equation}
 the integration constant $M$ is related to the mass of the BH. The most interesting feature of this solution is the constant potential $\zeta$, which reflects the presence of a global monopole. A global monopole is a type of topological defect that can form in the early universe due to the spontaneous breaking of a global $O(3)$ symmetry down to a global $O(2)$ symmetry \cite{r1b}. A global monopole has a spherical shape and a solid deficit angle in its asymptotic geometry. Usually, a global monopole does not couple to gravity, but in this solution, the global monopole is induced by the mass of the graviton, which modifies the Einstein equations \cite{r1a}. This solution is similar to the 4D solution found in Ref.\cite{r1a}, but with an extra dimension.\\
The BH`s entropy is
\begin{equation}
   S = \pi r_+^2 
\end{equation}
At the event horizon, the mass and temperature are computed as
\begin{equation}
    T = {\frac {-8\,P \left( 6\, \left( 1/3+2/3\,\alpha+\beta \right) c{
m_g}^{2}{r_+}^{3}+ \left( -1-3\, \left( \beta+\alpha/3 \right) {m_g}^{2}{c}^
{2} \right) {r_+}^{2}+{Q}^{2} \right) \pi +3\,{r_+}^{4}}{32{\pi }^{2}{r_+}^{3}
P}}
\end{equation}
\begin{equation}
    M={\frac{ 8\,P\pi( \alpha\,{c}^{2}{m_g}^{2}{r_+}^{2}-2 \,
\alpha\,c{m_g}^{2}{r_+}^{3}+3\,\beta\,{c}^{2}{m_g}^{2}{r_+}^{2}-3\beta\,c{m_g}^{2}{r_+}^{3}-c{m_g}^{2}{r_+}^{3}+{Q}^{
2}+{r_+}^{2})+{r_+}^{4}}{16 \pi \,Pr_+}}
\end{equation}
 \subsection{Topology}
\begin{figure}
    \centering
\includegraphics[width=8cm]{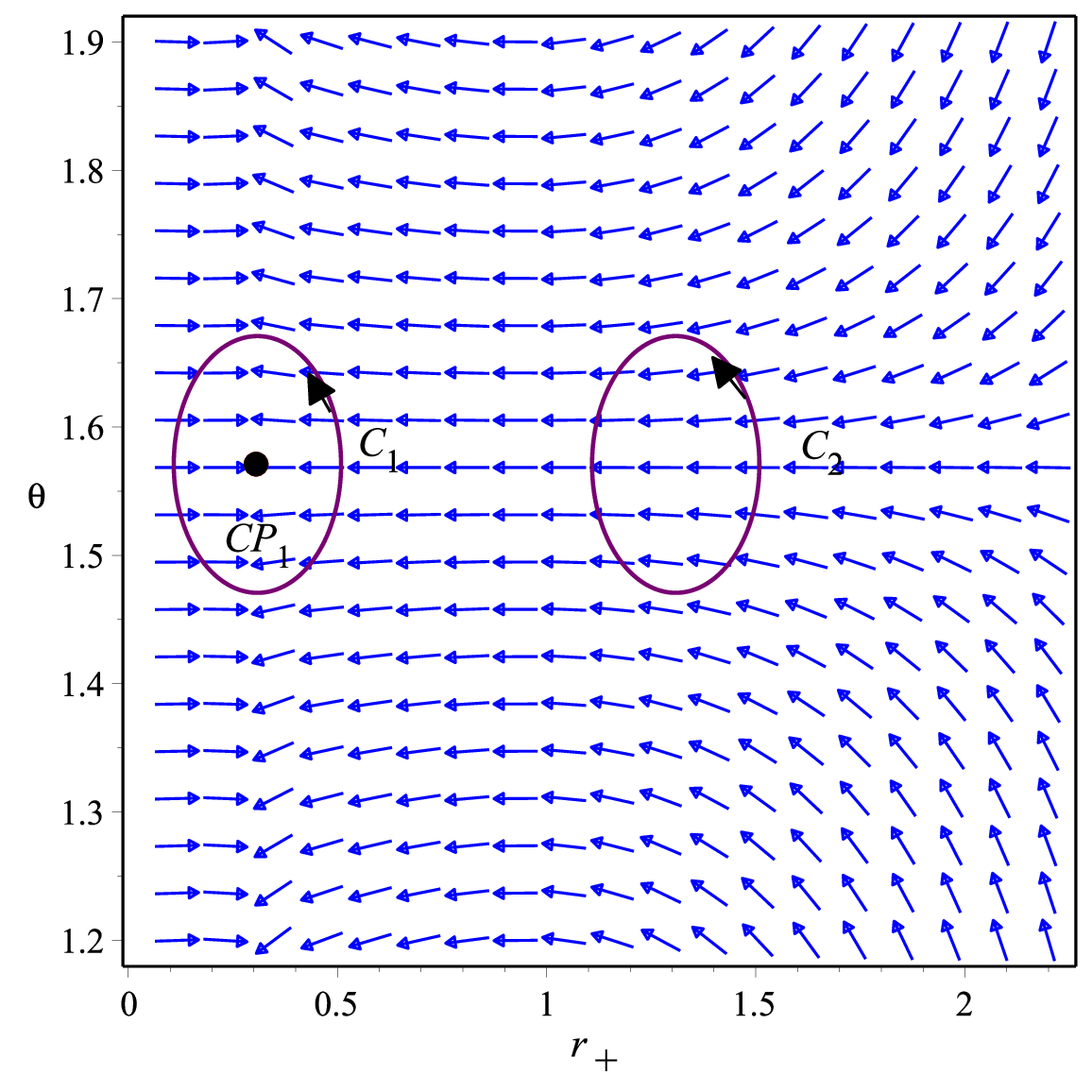}
\includegraphics[width=8cm]{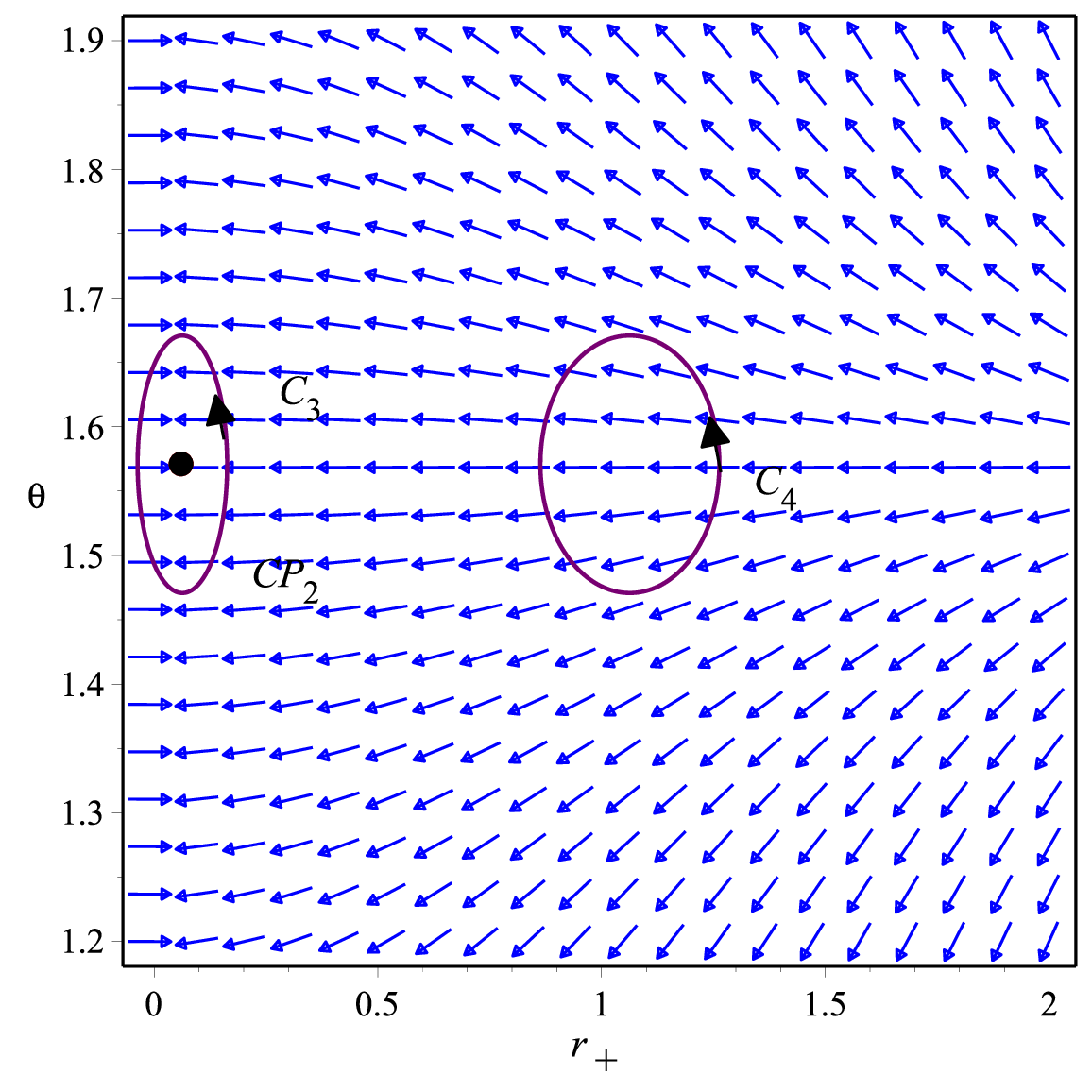}
    \caption{The blue arrows represent the vector field n of the $r_+-\theta$ plane for a charged BH in dRGT massive gravity, Left panel:$Q = c = m_g = 1$, $\alpha = 2$ and $\beta = 10$; Right panel: $Q = m_g = 1$, $c = 10$, $\alpha = -15$ and $\beta = 10$ .}
    \label{fig1}
\end{figure}

In order to examine the thermodynamic topology of BHs in dRGT massive gravity dimensions, the relationship between the charge, horizon radius, and pressure can be used to express the temperature.
\begin{equation}\label{13}
    T = \frac {-8\,P \left( 6\, \left( 1/3+2/3\,\alpha+\beta \right) c{
m}^{2}{r_+}^{3}+ \left( -1-3\, \left( \beta+\alpha/3 \right) {m}^{2}{c}^
{2} \right) {r_+}^{2}+{Q}^{2} \right) \pi +3\,{r_+}^{4}}{{32\pi }^{2}{r_+}^{3}
P}
\end{equation}
The associated critical points are established by 
\begin{equation}\label{14}
    \frac{\partial T}{\partial r_+} = 0, \frac{\partial^2T}{\partial^2r_+} = 0
\end{equation}
Use of the condition $(\frac{\partial_{r_+} T}{\partial_{r_+} S})_{Q, P} = 0$, this results in a formula for the pressure, P.
\begin{equation}\label{15}
    P=\frac {-3{r_+}^{4}}{8\pi \, \left( -\alpha\,{c}^{2}{m}^{2}{r_+}^{2}-
3\,\beta\,{c}^{2}{m}^{2}{r_+}^{2}+3\,{Q}^{2}-{r_+}^{2} \right) }
\end{equation}
Substituting P in eq.(\ref{13}), the term temperature, results in:
\begin{equation}\label{16}
   T =  \frac {-3\, \left( 1/3+2/3\,\alpha+\beta \right) c{m}^{2}{r_+}^{3}
+ \left( 1+3\, \left( \beta+\alpha/3 \right) {m}^{2}{c}^{2} \right) {r_+
}^{2}-2\,{Q}^{2}}{2{r_+}^{3}\pi }
\end{equation}
An updated thermodynamic function $\Phi$ is to be calculated as,
\begin{equation}\label{17}
    \Phi =  {\frac{1}{\sin{\theta}}}\,T(Q,r_+)=\frac {-3\, \left( 1/3+2/3\,\alpha+\beta \right) c{m}^{2}{r_+}^{3}
+ \left( 1+3\, \left( \beta+\alpha/3 \right) {m}^{2}{c}^{2} \right) {r_+
}^{2}-2\,{Q}^{2}}{2\sin{\theta}{r_+}^{3}\pi }
\end{equation}
The vector field`s $ \phi = (\phi^{r_+},\phi^\theta)$ components are
\begin{equation}\label{18}
  \phi^{r_+} = \frac { \left( -1-3\, \left( \beta+\alpha/3 \right) {m}^{2}{c}^{
2} \right) {r_+}^{2}+6\,{Q}^{2}}{2\sin \left( \theta \right) {r_+}^{4}\pi }
\end{equation}
and 
\begin{equation}\label{19}
    \phi^{\theta} = \frac { \left( 3/2\, \left( 1/3+2/3\,\alpha+\beta \right) c{m}^{2}{r_+}
^{3}+ \left( -1/2-3/2\, \left( \beta+\alpha/3 \right) {m}^{2}{c}^{2}
 \right) {r_+}^{2}+{Q}^{2} \right) \cos \left( \theta \right) }{ \left( 
\sin \left( \theta \right)  \right) ^{2}{r_+}^{3}\pi }
\end{equation}
The components of the normalized vector are
\begin{equation}\label{20}
    \frac{\phi^{r_+}}{\|\phi\|} =
\frac {3\left( -1/6- \left( \beta+\alpha/3 \right) 2{m}^{2}{c}
^{2} \right) {r_+}^{2}+{Q}^{2}}{{\sin \left( \theta \right) {r_+}^{4}\pi }\sqrt {{\frac { \left(  \left( -1-3\, \left( \beta+
\alpha/3 \right) {m}^{2}{c}^{2} \right) {r_+}^{2}+6\,{Q}^{2} \right) ^{2
}}{4 \left( \sin \left( \theta \right)  \right) ^{2}{r_+}^{8}{\pi }^{2}}}
+{\frac { \left( 3/2\, \left( 1/3+2/3\,\alpha+\beta \right) c{m}^{2}{r_+
}^{3}+ \left( -1/2-3/2\, \left( \beta+\alpha/3 \right) {m}^{2}{c}^{2}
 \right) {r_+}^{2}+{Q}^{2} \right) ^{2} \left( \cos \left( \theta
 \right)  \right) ^{2}}{ \left( \sin \left( \theta \right)  \right) ^{
4}{r_+}^{6}{\pi }^{2}}}}}
\end{equation}
and
\begin{equation}\label{21}
    \frac{\phi^{\theta}}{\|\phi\|} = \frac { \left( 3/2\, \left( 1/3+2/3\,\alpha+\beta \right) c{m}^{2}{r_+}
^{3}+ \left( -1/2-3/2\, \left( \beta+\alpha/3 \right) {m}^{2}{c}^{2}
 \right) {r_+}^{2}+{Q}^{2} \right) \cos \left( \theta \right) }{{ \left( 
\sin \left( \theta \right)  \right) ^{2}{r_+}^{3}\pi }\sqrt {{\frac { \left(  \left( -1-3\, \left( \beta+\alpha/3 \right) {m}^
{2}{c}^{2} \right) {r_+}^{2}+6\,{Q}^{2} \right) ^{2}}{ 4\left( \sin
 \left( \theta \right)  \right) ^{2}{r_+}^{8}{\pi }^{2}}}+{\frac {
 \left( 3/2\, \left( 1/3+2/3\,\alpha+\beta \right) c{m}^{2}{r_+}^{3}+
 \left( -1/2-3/2\, \left( \beta+\alpha/3 \right) {m}^{2}{c}^{2}
 \right) {r_+}^{2}+{Q}^{2} \right) ^{2} \left( \cos \left( \theta
 \right)  \right) ^{2}}{ \left( \sin \left( \theta \right)  \right) ^{
4}{r_+}^{6}{\pi }^{2}}}}}
\end{equation}
The normalized vector $n = (\frac{\phi^{r_+}}{\|\phi\|},\frac{\phi^{\theta}}{\|\phi\|})$ is expressed in Fig.(\ref{fig1}). The plot expresses the vector plot of $n$ in a $r_+$-$\theta$ plane for a BH in dRGT massive gravity. We have fixed $Q = c = m_g = 1$, $\alpha = 2$, $\beta = 10$ and $Q = m_g = 1$, $\alpha = -15$, $\beta = 10$, $c = 10$ for this plot respectively. The black dots represents the critical point $(CP_1)$ and $(CP_2)$, located at $(r_+,\theta) = (0.4264,\frac{\pi}{2})$ and $(0.063,\frac{\pi}{2})$, respectively. \\
The topological charge related
 with the critical point calculated as a contour $C$ is defined, parameterized by $\vartheta$ $\epsilon$ $(0,2\pi)$, as described in \cite{a8}.
\begin{equation}\label{22}
     r_+ = \alpha\cos{\vartheta}+ r_o,\,
     \theta = \beta\sin{\vartheta}+ \frac{\pi}{2}
    \end{equation}
     Next, $C_1$ \& $C_2$, two contours, are created such that $C_1$ encloses $CP_1$, while $C_2$ does not. For such contours we take $(\alpha,\beta,r_o)$ = $(0.2,0.1,0.4264)$ and $(0.2,0.1,1.4264)$ respectively.\\
    The deviation of the vector field $n$ as it moves in parallel with the contour $C$ can be characterized as:
   \begin{equation}\label{23}
       \Omega(\vartheta) = \int^\vartheta_0 \epsilon_{\alpha\beta} n^\alpha \partial_ \vartheta n^\beta d\vartheta.
   \end{equation}
 The topological charge is defined as $Q = \frac{1}{2\pi} \Omega(2\pi)$ where $\Omega(2\pi)$ is the solid angle subtended by a closed contour in the thermodynamic space. We consider two contours, $C_1$ and $C_2$, that enclose $CP_1$ and no critical point, respectively. We find that the topological charge of $CP_1$ is $Q_{CP_1}=$ $-1$, which indicates a conventional critical point. The topological charge of $C_2$ is zero. Therefore, the total topological charge is
\begin{equation}
Q = -1
\end{equation}
 \begin{figure}
    \centering
    \includegraphics[width=8cm]{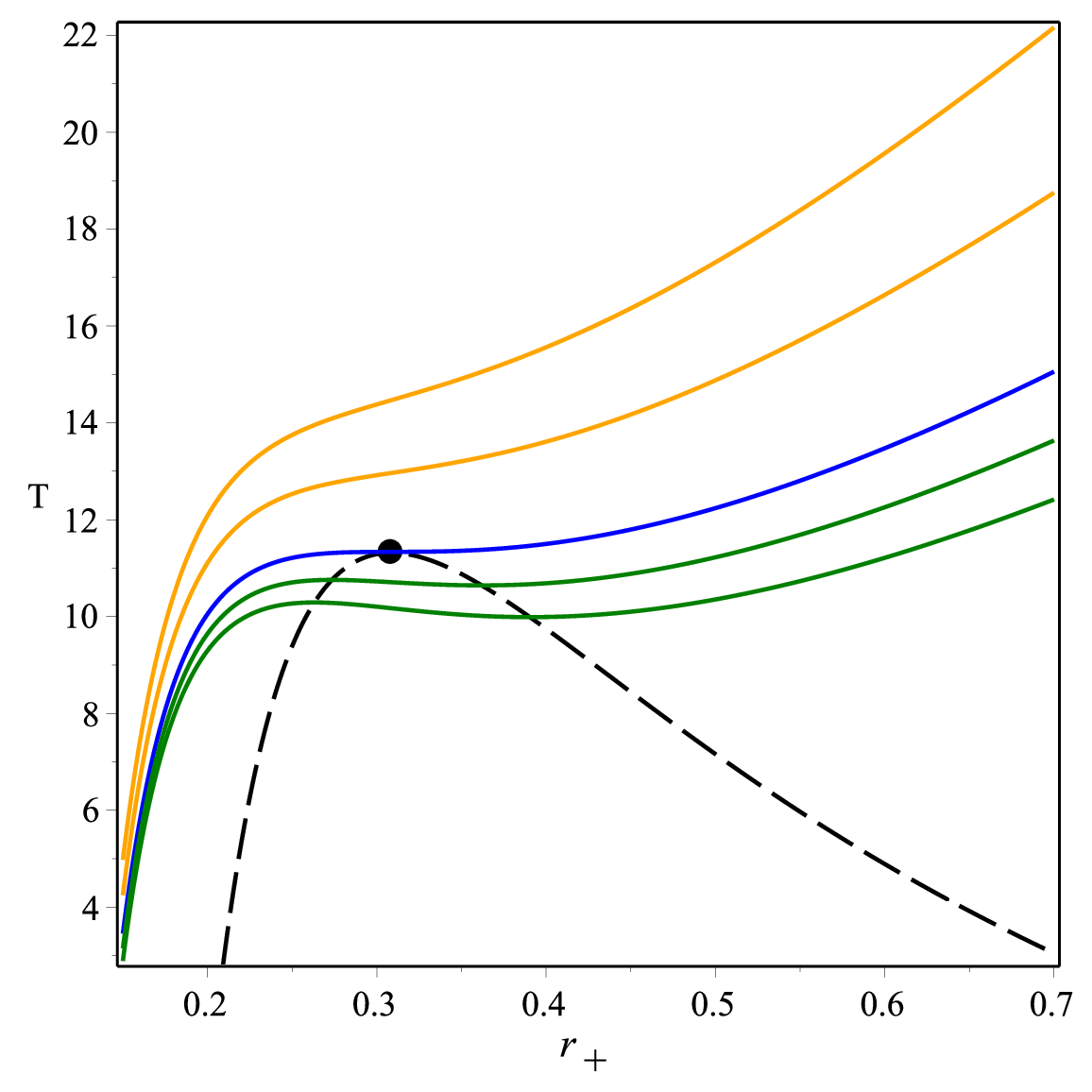}
    \caption{Isobaric curves (blue, green, and orange) for the charged BH in dRGT massive gravity for $Q = c = m_g = 1$, $\alpha = 2$ and $\beta = 10$ shown in $T-r_+$ plane. The black dot represents the critical point, and the black dashed curve is for the extremal points of the temperature.}
    \label{fig2}
\end{figure}
We also note that the critical radius given by Eq.(\ref{14}) matches exactly with the coordinates of $CP_1$, which are $(r_+,\theta)=(0.4264,\frac{\pi}{2})$. To analyze the properties of $CP_1$, we create a visual representation of the phase diagram (isobaric curves) around it in Fig.(\ref{fig2}), where we mark $CP_1$ by a black dot. The blue curve represents the isobaric curve for $P= P_c$, where $P_c$ is the critical pressure. The green curves represent the isobaric curves for $P>P_c$, and the orange curves represent the isobaric curves for $P<P_c$. We also use Eq.(\ref{16}) to plot the extremal points, depicted by the black dashed curve where the specific heat diverges. As shown in Figure.(\ref{fig2}), there is an unstable region at $P_c$ that separates the small and large BH phases, which can be seen by the region bounded by the two extremal points of each isobaric curve.
The presence of $CP_1$ leads to the disappearance of multiple phases of a BH in dRGT massive gravity, with one phase completely vanishing at $CP_1$. This means that $CP_1$ is also a phase annihilation point, where two phases merge into one.
 \subsection{Thermodynamic Topological Defects}
 The analysis now explores the BH solution in dRGT massive gravity\cite{r1} as topological thermodynamic defects. The expression of the generalized free energy can be derived by utilizing the mass and entropy of the BH.
 \begin{equation}\label{24}
     F = \frac {-16\,P{r_+}^{3}{\pi }^{2}+8\,Pt \left( -3\,c \left( \beta+
2/3\,\alpha+1/3 \right) {m}^{2}{r_+}^{3}+ \left( 1+3\, \left( \beta+
\alpha/3 \right) {m}^{2}{c}^{2} \right) {r_+}^{2}+{Q}^{2} \right) \pi +{
r_+}^{4}t}{16\pi \,Pr_+t}
 \end{equation}
 The components of the vector field, as given by Eq.(\ref{7}), are:
 \begin{equation}\label{25}
   \phi^{r_+} = \frac {-32\,P{r_+}^{3}{\pi }^{2}-8\, \left( 6\,c \left( \beta+2/3
\,\alpha+1/3 \right) {m}^{2}{r_+}^{3}+ \left( -1-3\, \left( \beta+\alpha
/3 \right) {m}^{2}{c}^{2} \right) {r_+}^{2}+{Q}^{2} \right) tP\pi +3\,{r_+
}^{4}t}{16P{r_+}^{2}t\pi }
\end{equation}
and
\begin{equation}\label{26}
    \phi^\Theta = -\cot\Theta \csc\Theta
\end{equation}
The unit vectors that corresponds are
\begin{equation}\label{27}
    n^1 = \frac {-4\,P{r_+}^{3}{\pi }^{2}+Pt \left( -4\, \left( \alpha+3/2\,
\beta+1/2 \right) c{m}^{2}{r_+}^{3}+ \left( 1+{m}^{2} \left( \alpha+3\,
\beta \right) {c}^{2} \right) {r_+}^{2}-{Q}^{2} \right) \pi +3/8\,{r_+}^{4
}t}{2{P\pi \,{r_+}^{2}t}\sqrt {{\frac { \left( -32\,P{r_+}^{3}{
\pi }^{2}-8\, \left( 6\,c \left( \beta+2/3\,\alpha+1/3 \right) {m}^{2}
{r_+}^{3}+ \left( -1-3\, \left( \beta+\alpha/3 \right) {m}^{2}{c}^{2}
 \right) {r_+}^{2}+{Q}^{2} \right) tP\pi +3\,{r_+}^{4}t \right) ^{2}}{256
\,{P}^{2}{r_+}^{4}{t}^{2}{\pi }^{2}}}+ \left( \cot \left( \Theta
 \right)  \right) ^{2} \left( \csc \left( \Theta \right)  \right) ^{2}
}}
\end{equation}
and
\begin{equation}\label{28}
    n^2  = \frac {-\cot \left( \Theta \right) \csc \left( \Theta \right)}{
\sqrt {{\frac { \left( -32\,P{r_+}^{3}{\pi }^{2}-8\, \left( 6\,c \left( 
\beta+2/3\,\alpha+1/3 \right) {m}^{2}{r_+}^{3}+ \left( -1-3\, \left( 
\beta+\alpha/3 \right) {m}^{2}{c}^{2} \right) {r_+}^{2}+{Q}^{2} \right) 
tP\pi +3\,{r_+}^{4}t \right) ^{2}}{256\,{P}^{2}{r_+}^{4}{t}^{2}{\pi }^{2}}
}+ \left( \cot \left( \Theta \right)  \right) ^{2} \left( \csc \left( 
\Theta \right)  \right) ^{2}}}
\end{equation}\\
\begin{figure}
    \centering
\includegraphics[width=8cm]{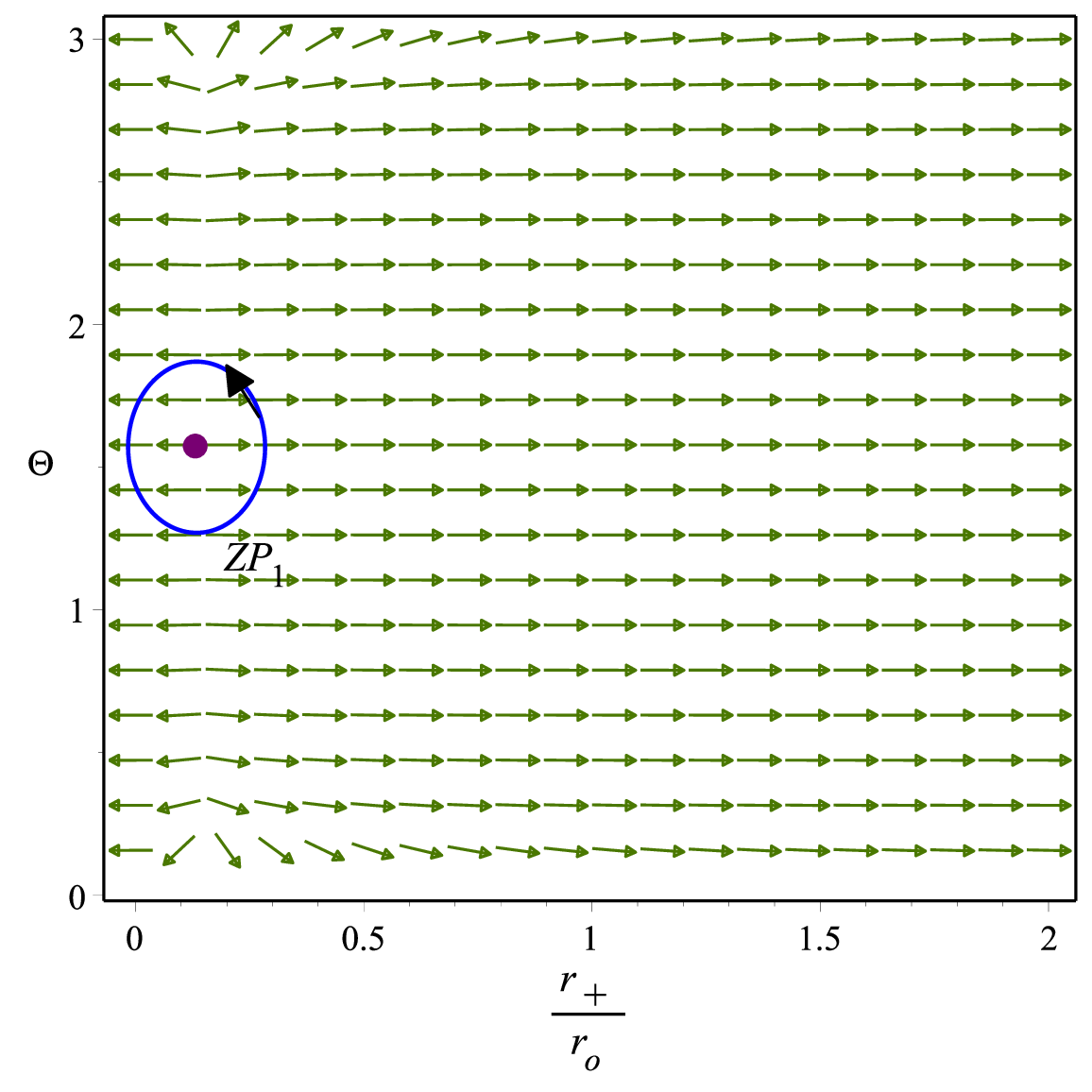}
    \caption{The green arrows represent the vector field n of the $r_+/r_o-\theta$ plane for a charged BH in dRGT massive gravity for P${r^2_o} = 0.0001$ (below the critical pressure $P_c$), $Q = c = m_g = 1$, $\alpha = 2$ and $\beta = 10$. The zero point is indicated by the purple dot.}
    \label{fig3}
\end{figure}
To determine the zero points, we plot and identify the unit vectors by putting $\Theta = \pi/2$ in $n^1$ for $\alpha = 2$, $\beta = 10$ and $c = 1$ 
and setting it equals to zero. For example, when $Q/r_o = 1$, $t/r_o = 4$, and $Pr^2_o = 0.0001$, we locate a zero point $(ZP_1)$ at $(r_+/r_o, \Theta) = (0.1497, \pi/2)$. The random length scale $r_o$ is ascertained by the dimensions of the cavity that encloses the BH, ensuring that P$<$ $P_c$. Figure (\ref{fig3}) illustrates the unit vectors and the zero point. The topological charge or winding number is \begin{equation}
    w=+1.
\end{equation}\\\\
\begin{figure}
    \centering
    \includegraphics[width=8cm]{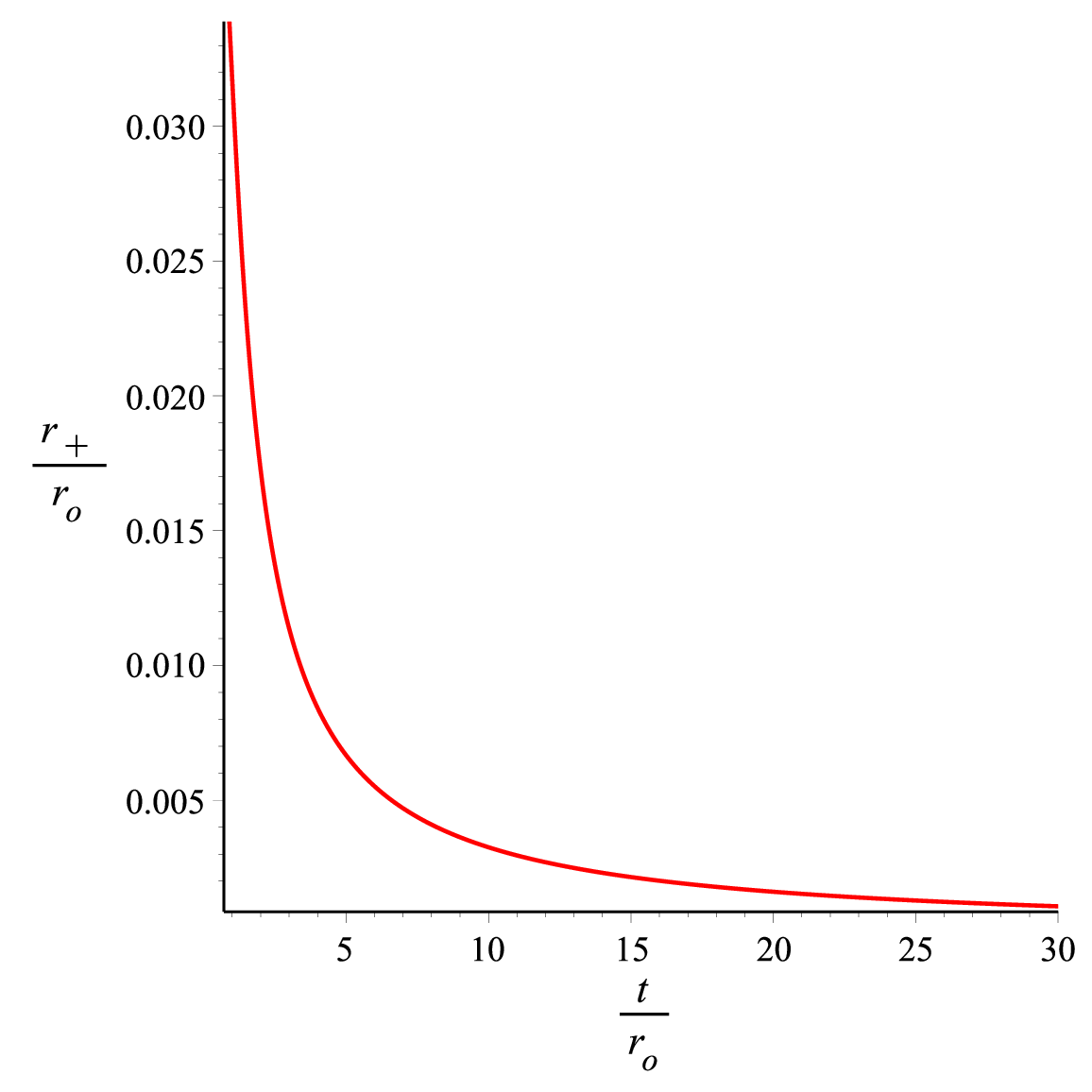}
    \caption{ The zero point of $\phi^{r_+}$ in $t/r_o$ vs. $r_+/r_o$ plane for a charged BH in dRGT massive gravity for P$<$ $P_c$}.
    \label{fig4}
\end{figure}
An analytical expression for $t$ can be derived for the zero points by equating $\phi^{r_+}$ to zero.
\begin{equation}\label{29}
    t=\frac {16P{r_+}^{3}{\pi }^{2}}{P \pi r^2_+(8\,\alpha\,{c}^{2}{m}^{2}-16 \,\alpha\,c{m}^{2}{r_+}+24\,\beta\,{c}^{2}{m}^{2
}-24\,\beta\,c{m}^{2}{r_+}-8\,c{m}^{2}{r_+}+8 \,{Q}^{2}+8)+{r_+}^{4}}
\end{equation}\\
Figure. (\ref{fig4}) exhibits a graph of $r_+$ plotted against $t$, which was previously obtained. The curve's points represent the zeros of $\phi^{r_+}=0$. To generate this plot, we set $Q/r_o = 1$, and $Pr^2_o = 0.0001$ (P$<$ $P_c$). The plot depicts that there is only one thermodynamically stable BH in dRGT massive gravity irrespective of the value of $t$.\\\\\\ 
To determine the zero points, we plot and identify the unit vectors by putiing $\Theta = \pi/2$ in $n^1$ for $\alpha = -15$, $\beta = 10$ and $c = 10$ and setting it equals to zero. For example, when $Q/r_o = 1$, $t/r_o = 0.001$, and $Pr^2_o = 0.000001$, we locate the zero points $(ZP_2)$, and $(ZP_3)$, having winding number $+1, -1$, respectively. Figure. (\ref{fig4a}) represents the unit vectors and the zero points for a class of BH in dRGT massive gravity.. Therefore, the zero point's total topological charge or  number is
\begin{equation}
    w=+1-1 = 0.
\end{equation}
\begin{figure}
    \centering
    \includegraphics[width=8cm]{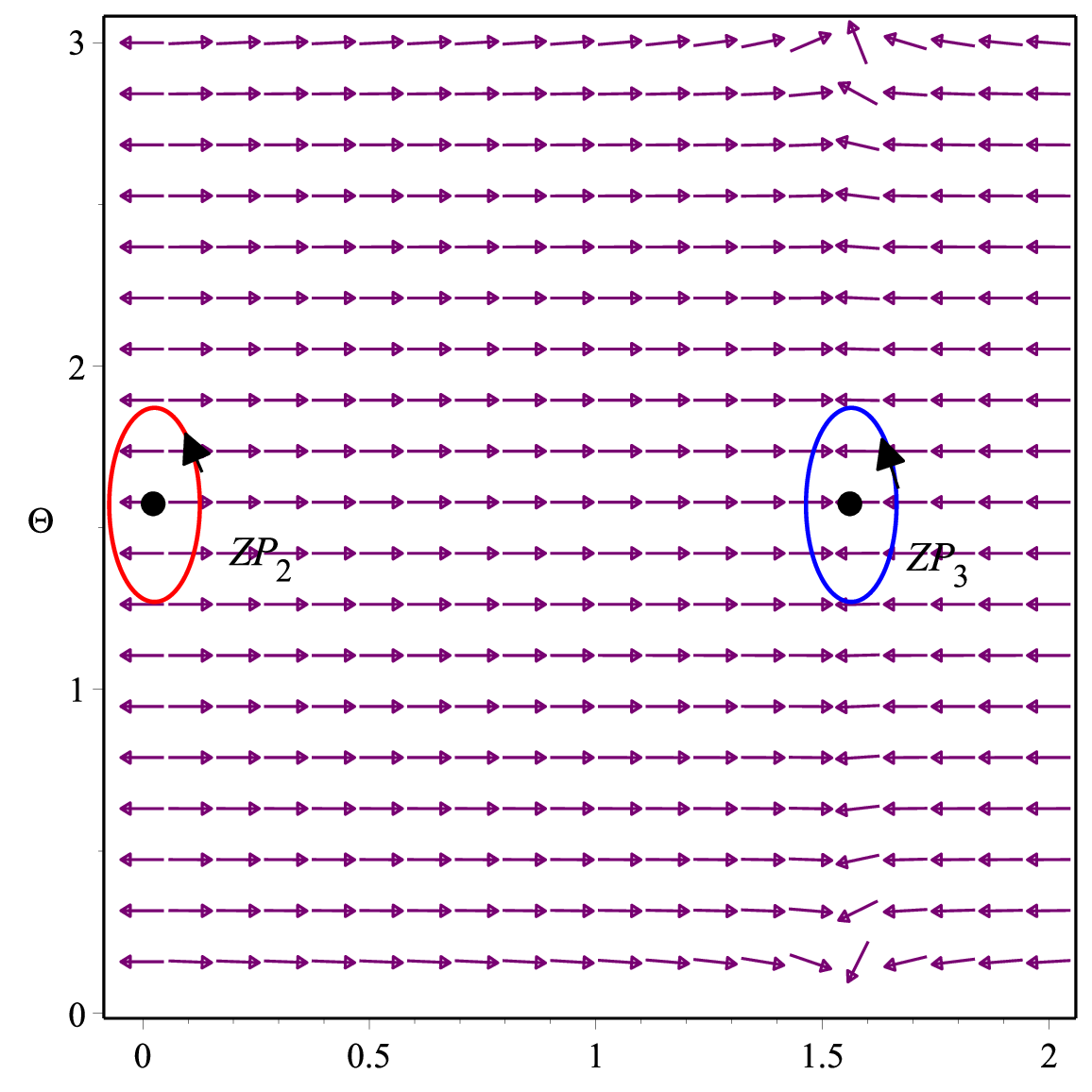}
     \caption{ The zero point of $\phi^{r_+}$ in $t/r_o$ vs. $r_+/r_o$ plane for a charged BH in dRGT massive gravity for $t/r_o = 2$}.
    \label{fig4a}
\end{figure}
\begin{figure}
    \centering
    \includegraphics[width=8cm]{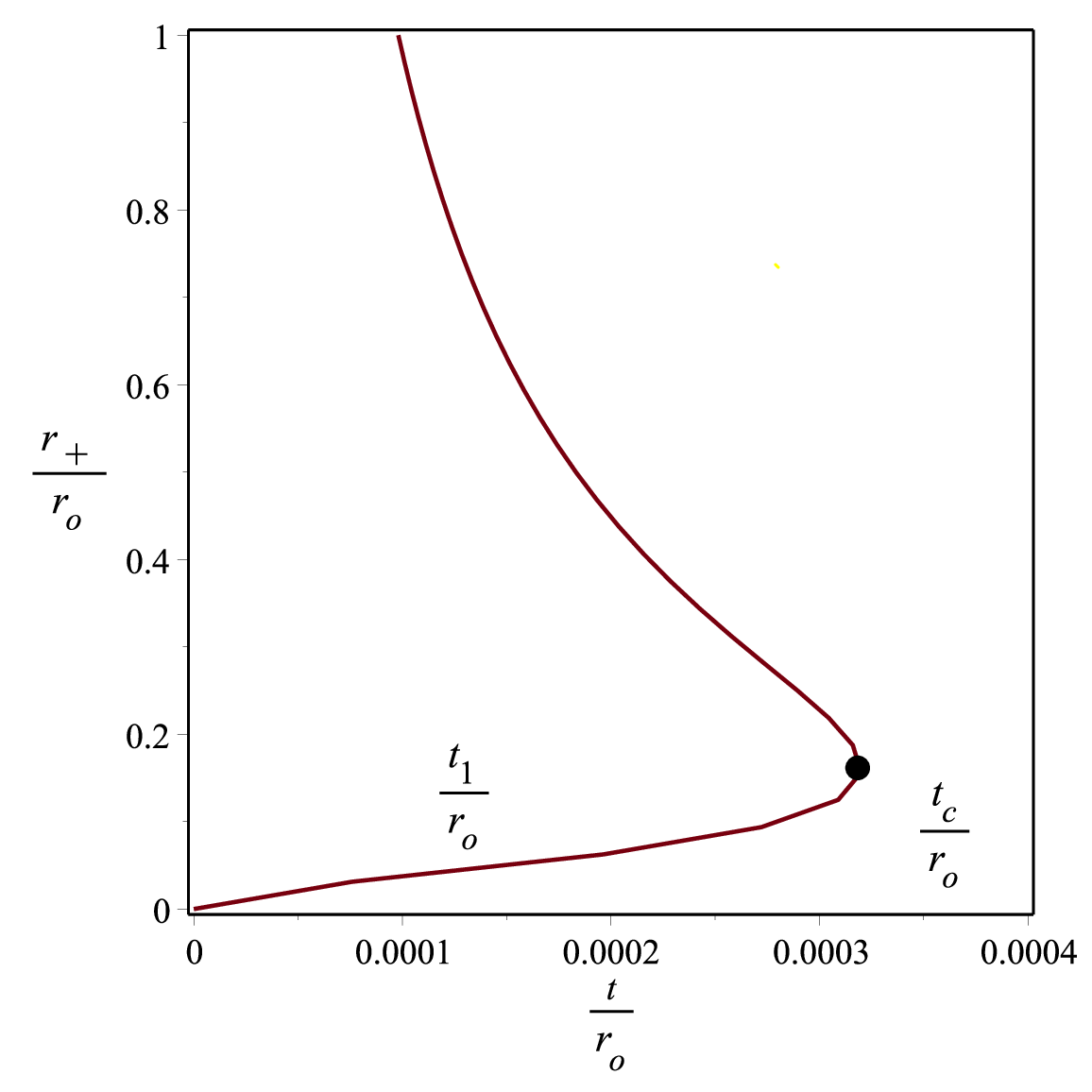}
    \caption{ The zero points of $\phi^{r_+}$ in $t/r_o$ vs. $r_+/r_o$ plane for a charged BH in dRGT massive gravity}.
    \label{fig4b}
\end{figure}
Figure. (\ref{fig4b}) exhibits a graph of $r_+$ plotted against $t$, which was previously obtained. The curve's points represent the zeros of $\phi^{r_+}=0$. To generate this plot, we set $Q/r_o = 1$, and $Pr^2_o = 0.000001$ (P$<$ $P_c$). The plot depicts that there is one stable and unstable BH in dRGT massive gravity. Eventually, the generation and annihilation points are identified by applying the condition $\partial_{r_+}F =\partial_{r_+,r_+}F = 0$. For $Q/r_o = 1$, and $Pr^2_o = 0.000001$.The generation and annihilation points are obtained
 at $t/r_o$ =  $t_1/r_o = 0.00032 $  which is shown as black dot in Fig.(\ref{fig4b}).\\\\\\
\section{Black hole solution in 5D Yang-Mills massive gravity}
The solution pertains to the five-dimensional planar AdS black brane presented in \cite{r2}.
\begin{equation}\label{e1}
    ds^2 = -f(r)dt^2 + \frac{{dr^2}}{{f(r)}} + r^2 h_{ij} dx^i dx^j
\end{equation}
where $h_{ij} = \frac{1}{b^2}\,\delta_{ij}$ and the constant "b" is an arbitrary length parameter. In spite of the fact that the presence of "b" has no effect on the spacetime geometry and can be adjusted to 1 without losing generality, it is required for dimensional analysis since we believe that $dx^i$ represents a length element. Alternatively, one can regard $dx^i$ as an element without dimensions, eliminating the need for "b".The reference metric ansatz we adopt is given by $f_{\mu\nu} = (0, 0, c_o^2,h_{ij})$, where $c_o$ is a positive constant. To simplify notation, we introduce the notation $c_i = m^2c_o^i,c_i$ for i = 1, 2, 3, representing the blackening factor associated with AdS radius $l = \sqrt{-\frac{6}{\lambda}}$  \cite{r2}
\begin{equation}\label{e2}
    f(r) = \frac{r^2}{l^2} - \frac{m_o}{r^2} + \frac{c_1}{3}r + c_2 + \frac{2c_3}{r} - \frac{2e^2}{r^2} \ln\left(\frac{r}{L}\right).
\end{equation}
in which the integration constant is denoted as “$m_o + 2e^2\ln(L)$”. It is important to emphasize that calculating gauge potentials for planar BHs is a challenging task. However, we can take advantage of the fact that the gauge field $F^{(a)}_{\mu\nu}$ and its associated energy-momentum tensor are unaffected by the horizon topology. Consequently, we can determine the metric function by examining the solutions of topological BHs with $(k = \pm 1)$ in the existence of the Wu-Yang ansatz. Notably, the parameter $m_o$ is linked to the BH's mass, while the term $2e^2\ln(L)$ is important to introduce a logarithmic term that does not have any dimensions with a random length factor L. Furthermore, it is important to point out that even if $c_o$ and b are two separate random constants with length dimensions, they can be set as $c_o$ = b or even $c_o$ = b = 1 without any loss of generality. This is because they lack physical characterization and do not affect the overall analysis.\\\\
By utilizing the fact that $f(r = r^+) = 0$, the integration constant can be eliminated, resulting in a rewritten expression of Eq.(\ref{e2}) as
 \begin{equation}\label{30e}
    f(r) =\frac{1}{r^2}\,\left({\frac {{r}^{4}-r_+^4}{{l}^{2}}}+\frac{c_1}{3}({r}^{3}-r^3_+)+c_2({r}^{2}-r^2_+)+2\,c_3(r-r_+)-2\,{e}^{2}\ln 
 \left( \frac{r}{r_+} \right)\right)  
 \end{equation}
 As anticipated, when the variables for massive and Yang-Mills terms approach zero, the metric function takes on a well-known form of $f(r) = \frac{r^2}{l^2} - \frac{r_+^4}{r^2 l^2}$ for the 5D planar AdS spacetime. By examining Eq.(\ref{e1}), we can calculate the Kretschmann scalar.
 \begin{equation}\label{e3}
     R_{\alpha\beta\gamma\delta} R^{\alpha\beta\gamma\delta} = \left(\frac{{d^2f(r)}}{{dr^2}}\right)^2 + 6\left(\frac{1}{r}\frac{{df(r)}}{{dr}}\right)^2 + 12\left(\frac{{f(r)}}{{r^2}}\right)^2
 \end{equation}
Upon utilizing Eq.(\ref{30e}), it is determined that a curvature singularity emerges at $r = 0$ and is concealed by an event horizon located at $r_+$. Consequently, the obtained solution can be interpreted as a planar black brane. 
The dominant term in Eq.(\ref{30e}) for large values of $r$ indicates the asymptotic behavior of the solutions, which aligns with the expected AdS characteristics. Furthermore, as $r$ approaches infinity, it is observed that the Kretschmann scalar tends towards $\frac{40}{l^2}$ $(R_{\alpha\beta\gamma\delta} R^{\alpha\beta\gamma\delta} \rightarrow \frac{40}{l^2})$, providing potential confirmation of the asymptotic AdS property of the solutions.
Nonetheless, It is important to check that the asymptotic symmetry group is the same as the AdS group in order to guarantee the existence of asymptotically AdS solutions.
We then calculate the thermodynamic quantities associated with this solution. We supply the results of the conservation and thermodynamic quantities per unit dimensionless volume by writing $\Omega_3 = \frac{V_3}{b^3}$, where $V_3$ is the volume of the hypersurface with radii $t$ and $r_+$ that is constant in both $t$ and $r$. We begin by calculating the entropy of the answer. Since we are operating under the premise of Einstein gravity, we use the Bekenstein-Hawking area law to do this. The area of the horizon is expressed as
\begin{equation}\label{e4}
    A = \int d^3x \sqrt{-g} \bigg|_{r=r+,t=cte} = \frac{V_3}{b^3} r_+^3
\end{equation}
Therefore, the entropy per unit volume $V_3$ can be expressed as
 \begin{equation}\label{31}
 S = \frac{A}{4} = \frac{r_+^3}{4}.
     \end{equation}
     For the metric of the form (\ref{e1}),the Hawking temperature can be calculated as
     \begin{equation}\label{e5}
         T = \frac{1}{4\pi}\frac{d f(r)}{dr}\bigg|_{r=r+} (9) = \frac{1}{2\pi r_+^3}\left(\frac{2r_+^4}{l^2} + \frac{1}{2} c_1r_+^3 + c_2r_+^2 + c_3r_+ - e^2)\right.
     \end{equation}
   The quantity $\frac{-16\pi G_d}{(d-2)\Omega_{(d-2)}}\frac{M}{r^{d-3}}$is extracted from the blackening factor (\ref{30e}) and used to compute the ADM mass of the black brane in $d$ dimensions. For the sake of simplicity, we'll assume that $G_5 = 1$ and write $\Omega_3 = \frac{V_3}{b^3}$ to represent the volume of the $(d-2)$-dimensional unit hypersurface. Therefore, the mass term per unit volume $V_3$ can be written as
     \begin{equation}\label{e6}
         M = \frac{3m_0}{16\pi} = \frac{3}{16\pi}\left({\frac {r_+^4}{{l}^{2}}}+\frac{c_1}{3}(r^3_+)+c_2(r^2_+)+2\,c_3(r_+)-2\,{e}^{2}\ln \left(\frac{r_+}{L}\right) \right)
     \end{equation}
    Here, $m_0$ is determined by setting the metric function (\ref{e2}) to zero at the event horizon $r_+$. The study of topological thermodynamics for a BH in Yang-Mills massive gravity has been previously addressed in \cite{r4}.
 \subsection{Thermodynamic Topological Defects}
\begin{figure}
    \centering
    \includegraphics[width=8cm]{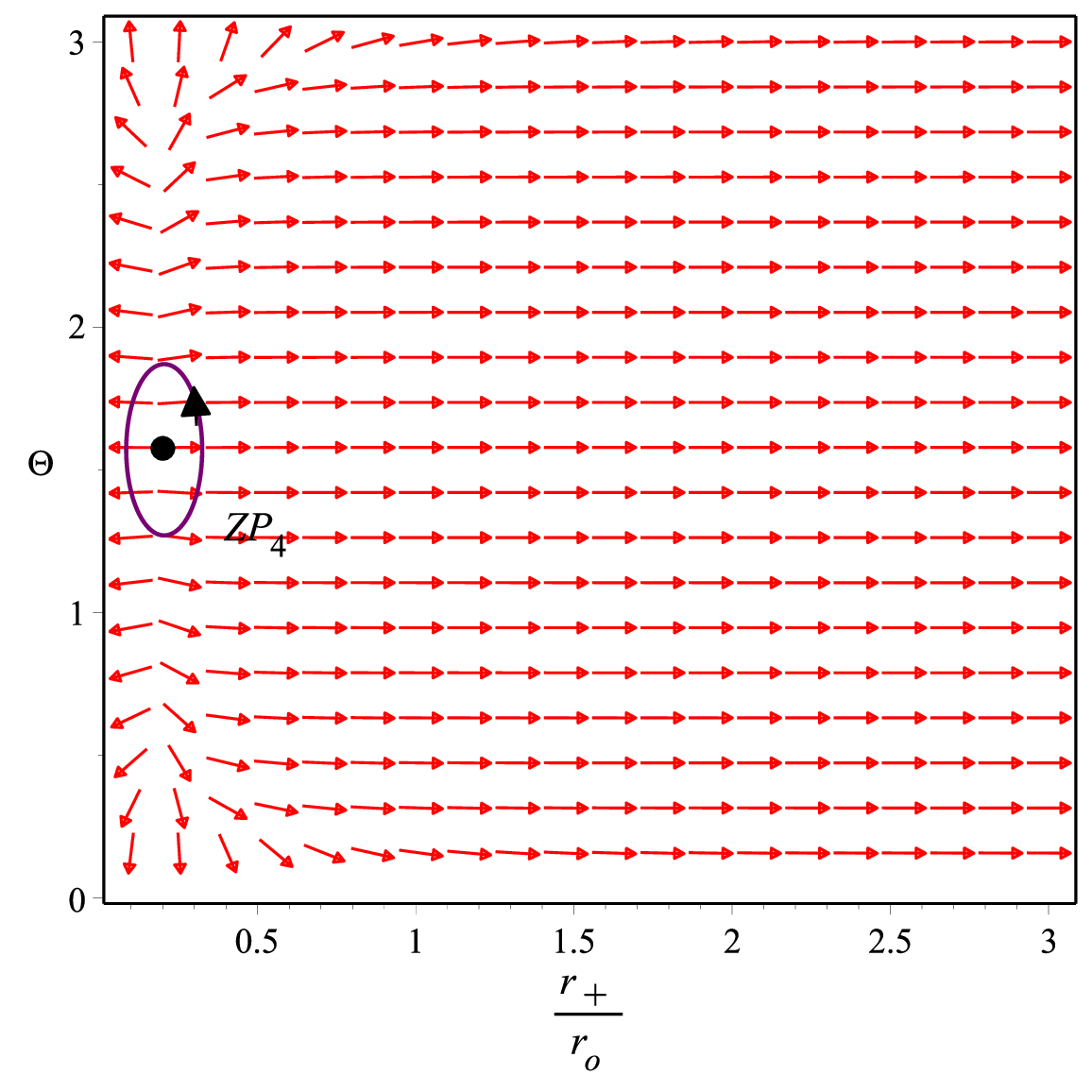}
    \caption{The red arrows represent the vector field n of the $r_+/r_o-\Theta$ plane for a BH in 5D Yang-Mills massive gravity for P${r^2_o} = 0.0004$ (P$<$ $P_c$). The zero point is indicated by the black dot.}
    \label{fig6}
\end{figure}

The present investigation delves into the examination of the topological thermodynamic defects of a BH solution in 5D Yang-Mills massive gravity \cite{r2}. By employing the mass and entropy of the BH, the generalized free energy can be formulated as
 \begin{equation}\label{43}
     F = {\frac {3}{16\,\pi } \left( {\frac {4{r_+}^{4}}{3\pi \,P}}+\frac{c_1{r_+}^{3}}{3}
+c_2{r_+}^{2}+2\,c_3r_+-2\,{e}^{2}\ln  \left( {\frac {r_+}{L}} \right)  \right) 
}-{\frac {{r_+}^{3}}{4t}}
 \end{equation}
 The components of the vector field, as given by Eq.(\ref{6}), are:
 \begin{equation}\label{44}
   \phi^{r_+} = -{\frac {-3\,P\pi \,c_1{r_+}^{3}t+12\,{\pi }^{2}P{r_+}^{3}-6\,P\pi \,c_2
{r_+}^{2}t-6\,P\pi \,c_3r_+t+6\,P\pi \,{e}^{2}t-16\,{r_+}^{4}t}{16\,{\pi }^{2}Pr_+t}
}
\end{equation}
and
\begin{equation}\label{45}
    \phi^\Theta = -\cot\Theta \csc\Theta
\end{equation}
The unit vectors that corresponds are as follows:
\begin{equation}\label{46}
    n^1 = -{\frac {-3\,P\pi \,c_1{r_+}^{3}t+12\,{\pi }^{2}P{r_+}^{3}-6\,P\pi \,c_2
{r_+}^{2}t-6\,P\pi \,c_3r_+t+6\,P\pi \,{e}^{2}t-16\,{r_+}^{4}t}{16{\pi }^{2}Pr_+t \sqrt {{\frac { \left( -3\,P\pi \,c_1{r_+}^{3}t+12\,{\pi }^{2}P
{r_+}^{3}-6\,P\pi \,c_2{r_+}^{2}t-6\,P\pi \,c_3r_+t+6\,P\pi \,{e}^{2}t-16\,{r_+}^{
4}t \right) ^{2}}{256\,{\pi }^{4}{P}^{2}{r_+}^{2}{t}^{2}}}+ \left( \cot
 \left( \Theta \right)  \right) ^{2} \left( \csc \left( \Theta
 \right)  \right) ^{2}}}}
\end{equation}
and
\begin{equation}\label{47}
    n^2  = \frac{-\cot \left( \Theta \right) \csc \left( \Theta \right)}{
\sqrt {{\frac { \left( -3\,P\pi \,c_1{r_+}^{3}t+12\,{\pi }^{2}P{r_+}^{3}-6\,
P\pi \,c_2{r_+}^{2}t-6\,P\pi \,c_3r_+t+6\,P\pi \,{e}^{2}t-16\,{r_+}^{4}t
 \right) ^{2}}{256\,{\pi }^{4}{P}^{2}{r_+}^{2}{t}^{2}}}+ \left( \cot
 \left( \Theta \right)  \right) ^{2} \left( \csc \left( \Theta
 \right)  \right) ^{2}}}
\end{equation}\\\\\\
To determine the zero points, we plot and identify the unit vectors by setting $\Theta = \pi/2$ in $n^1$ for  $c_1 = c_2 = c_3 = 1$, and $e = 2$
and equating it to zero. For example, for $t/r_o = 20$, and $Pr^2_o = 0.0004$, we locate a zero point $(ZP_2)$ at $(r_+/r_o, \Theta) = (0.205, \pi/2)$. The magnitude of $r_o$ represents a random length scale determined by the dimensions of the cavity that encloses BH, while ensuring that the pressure remains below the critical pressure $P_c$.
Figure. (\ref{fig6}) represents the unit vectors and the zero point. The zero point's topological charge or winding number is
\begin{equation}
    w = 1
\end{equation}
\\\\
\begin{figure}
    \centering
    \includegraphics[width=8cm]{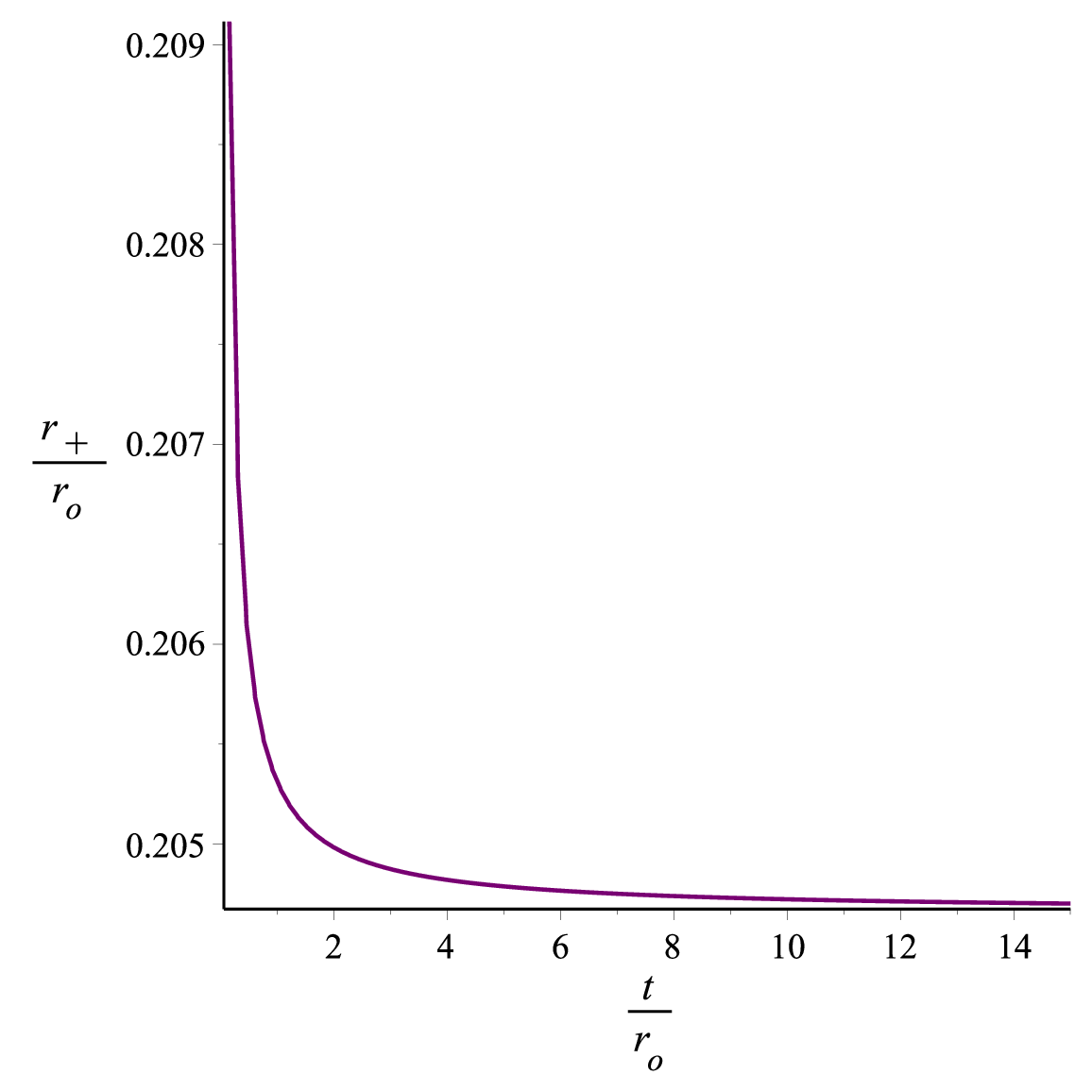}
    \caption{The zero point of $\phi^{r_+}$ in $t/r_o$ vs. $r_+/r_o$ plane for a charged BH in 5D Yang-Mills massive gravity for P$<$ $P_c$}.
    \label{fig7}
\end{figure}
An analytical expression for $t$ can be derived for the zero points by equating $\phi^{r_+}$ to zero.
\begin{equation}\label{48}
    t={\frac {12{\pi }^{2}P{r_+}^{3}}{3\,P\pi \,c_1{r_+}^{3}+6\,P\pi \,c_2{r_+}^{2
}+6\,P\pi \,c_3r_+-6\,P\pi \,{e}^{2}+16\,{r_+}^{4}}}
\end{equation}
Figure. (\ref{fig7}) exhibits a graph of $r_+$ plotted against $T$, which was previously obtained. The curve's points represent the zeros of $\phi^{r_+}=0$. To generate this plot, we set $Pr^2_o = 0.0004$ (P$<$ $P_c$). The plot depicts that there is only one thermodynamically stable BH in 5D Yang-Mills massive gravity irrespective of the value of $t$.\\\\\\
\begin{figure}
    \centering
    \includegraphics[width=8cm]{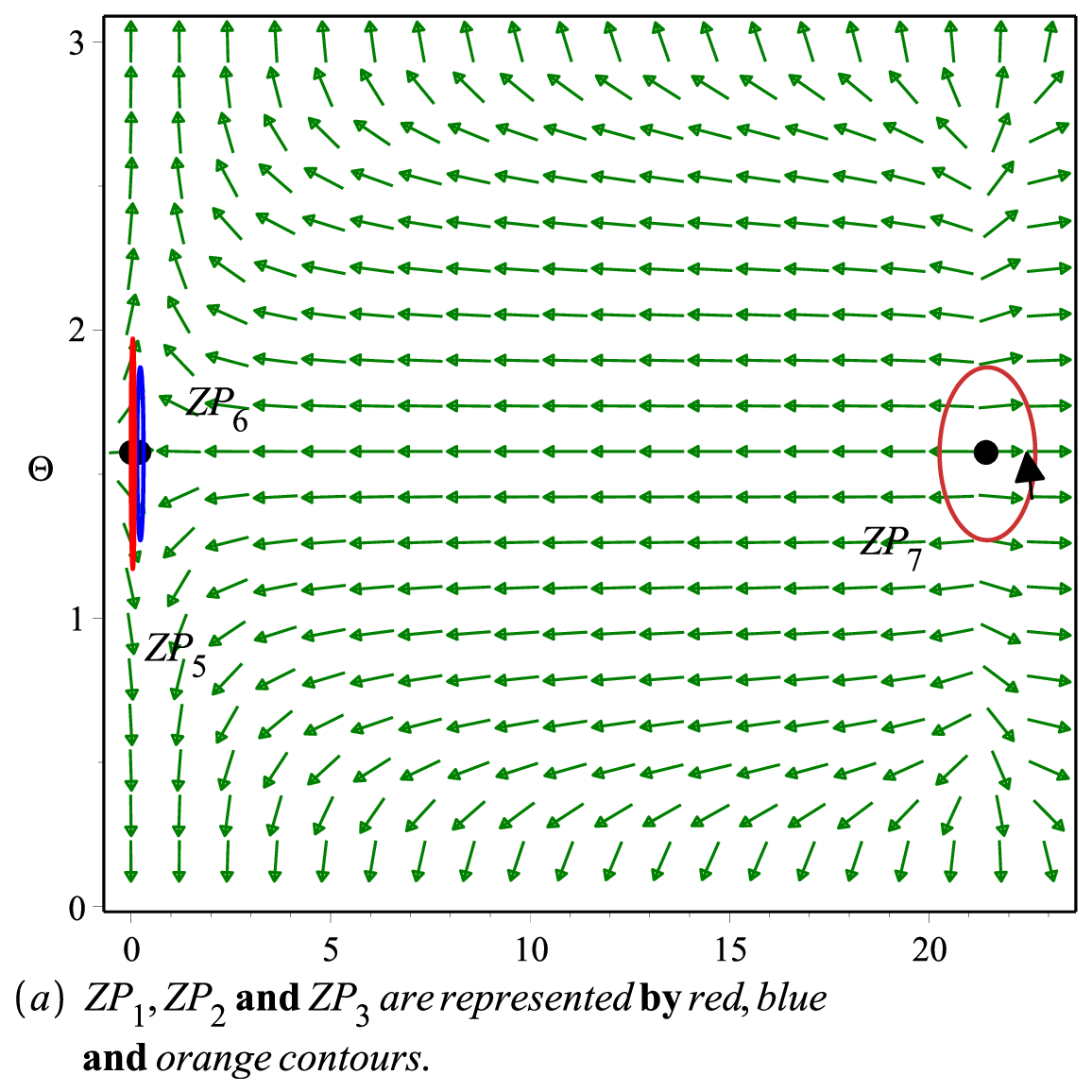}
    \includegraphics[width=8cm]{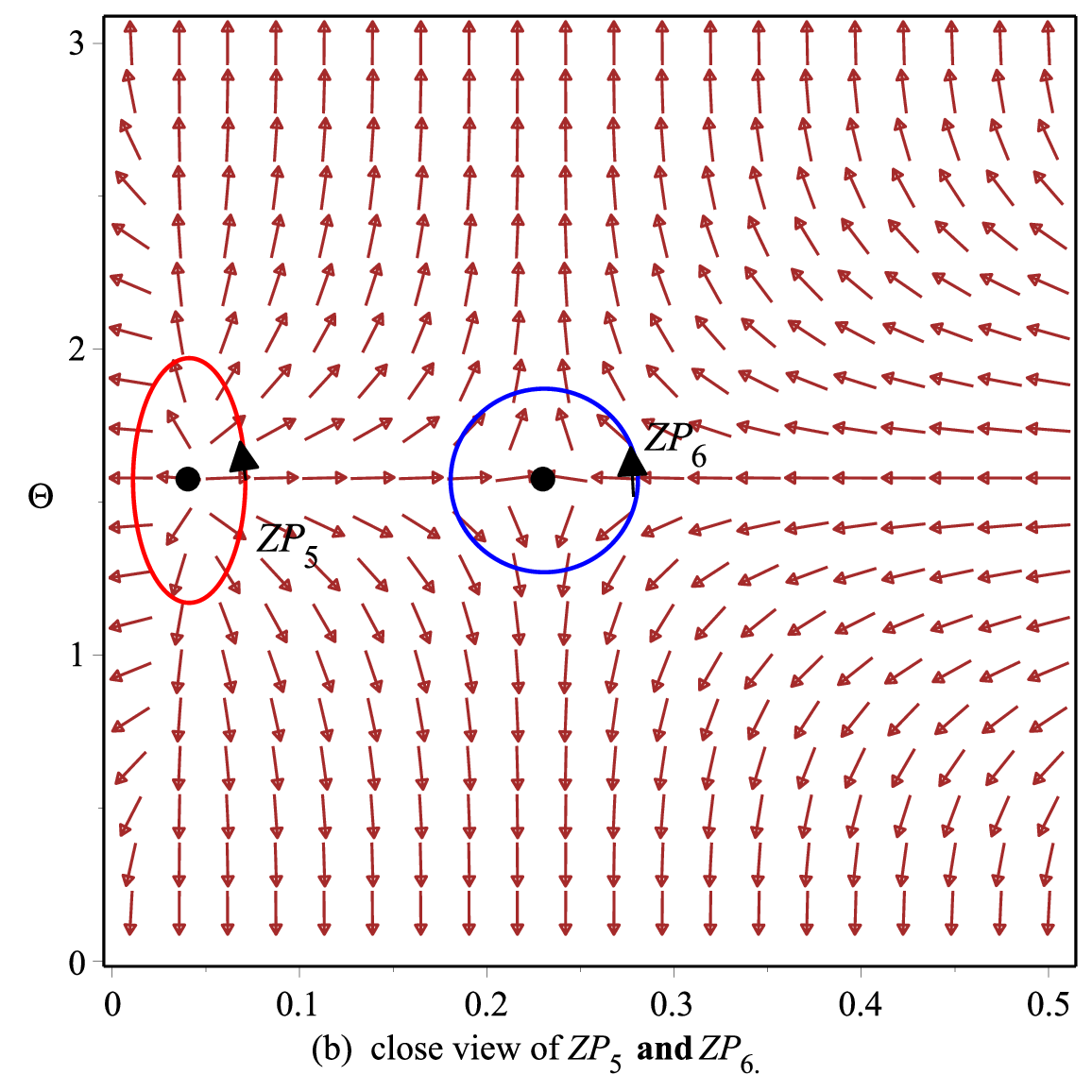}
    \caption{The green arrows represent the vector field n of the $r_+/r_o-\Theta$ plane for a BH in 5D Yang-Mills massive gravity for P${r^2_o} = 1.16$ (P$<$ $P_c$). The zero points are indicated by the black dot.}
    \label{fig6a}
\end{figure}
To determine the zero points, we plot and identify the unit vectors by putting $\Theta = \pi/2$ in $n^1$ for  $c_1 = 0.1, c_2 = -0.01, c_3 = 1$, and $e = 0.2$
and equating it equals to zero. For example, for $t/r_o = .4$, and $Pr^2_o = 1.16$, we locate zero points $(ZP_5)$, $(ZP_6)$ and $(ZP_7)$ at $(r_+/r_o, \Theta) = (0.04, \pi/2)$, $(0.23, \pi/2)$ and $(21.46, \pi/2)$. The magnitude of $r_o$ represents a random length scale determined by the dimensions of the cavity that encloses the BH, while ensuring that the pressure must be less than the critical pressure $P_c$. Figure. (\ref{fig6a}) represents the unit vector and the zero points with winding number $+1$, $-1$ and $+1$, respectively. Therefore, the zero point's total topological charge or winding number is
\begin{equation}
w = +1-1+1 = 1
\end{equation}
\\\\
\begin{figure}
    \centering
    \includegraphics[width=8cm]{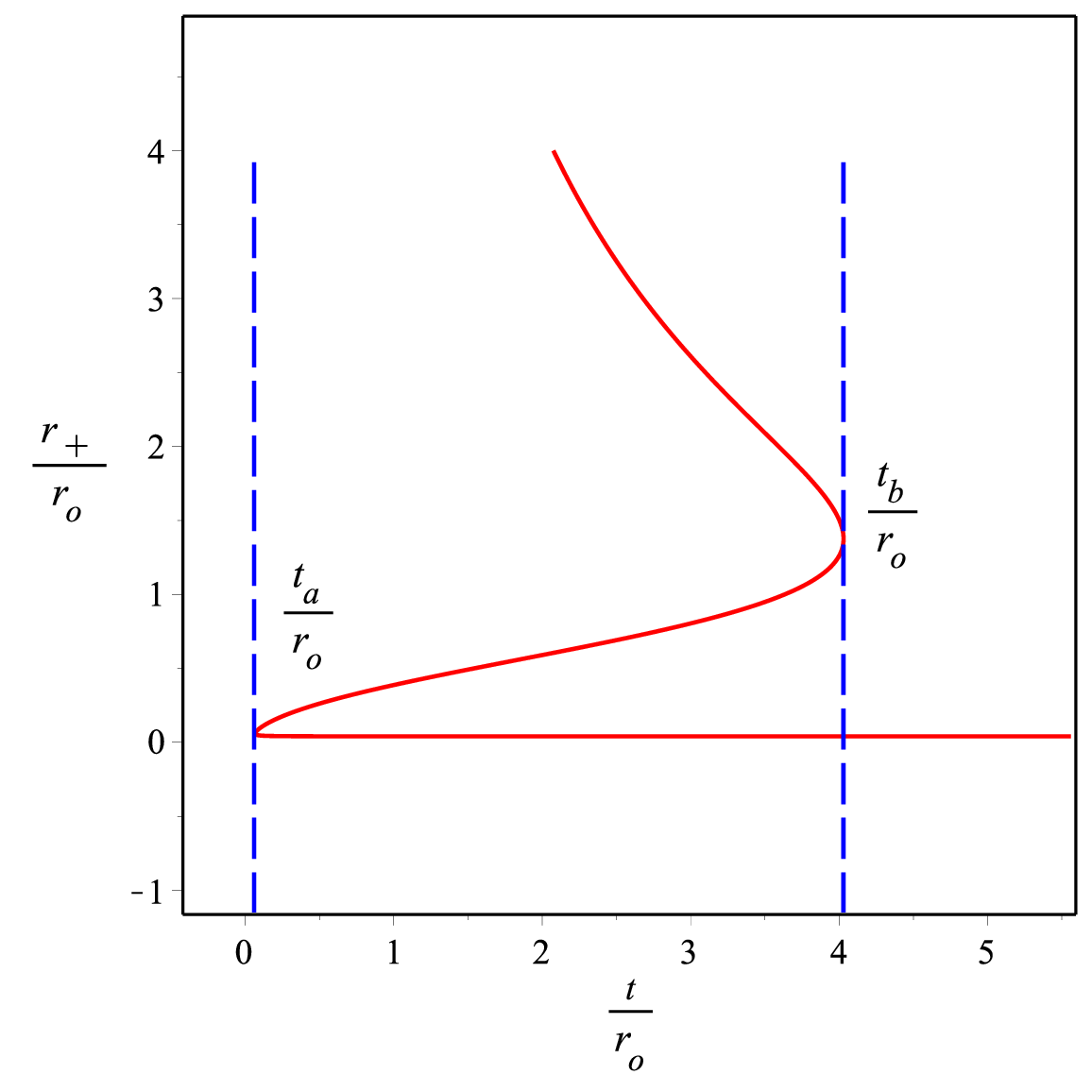}
    \caption{The zero points of $\phi^{r_+}$ in $t/r_o$ vs. $r_+/r_o$ plane for a charged BH in 5D Yang-Mills massive gravity for P$<$ $P_c$}.
    \label{fig7a}
\end{figure}
Figure. (\ref{fig7a}) exhibits a graph of $r_+$ plotted against $t$, which was previously obtained. The curve's points represent the zeros of $\phi^{r_+}=0$. To generate this plot, we set $Q/r_o = 1$, and $Pr^2_o = 1.16$ (P$<$ $P_c$). The plot shows that there is one stable and unstable BH in 5D Yang-Mills massive gravity.
Eventually, the generation and annihilation points are identified by applying the condition $\partial_{r_+}F =\partial_{r_+,r_+}F = 0$, which is shown by blue dashed lines in Fig.(\ref{fig7a}).\\\\\\
\section{ QUINTESSENCE SURROUNDING D-DIMENSIONAL RN-ADS BLACK
HOLES WITH A CLOUD OF STRINGS
}
Metrics describing d-dimensional spacetimes that are asymptotically AdS have recently been considered. This metric, together with its accompanying cloud of strings and quintessence, is produced by a charged static BH. According to \cite{r5}, it is supposed that the cloud of strings and quintessence have no interaction. The energy momentum tensor associated with the two sources is linearly combined in this case. The solution is described in its general version in \cite{r6}; this corresponds to a BH submerged in quintessence and interacting with a cloud of strings in a d-dimensional spacetime.
\begin{equation}\label{33}
    dS^2_d = -f(r)dt^2 + f(r)^{-1}dr^2 + r^2 d\Omega^2_{d-2}
\end{equation}
where, the metric on the unit $(d-2)$ sphere, denoted by $d\Omega^2_{d-2}$, is used to eliminate $\rho_q$ in the Einstein equation. By utilizing the metric ansatz mentioned above and the Einstein equation, we can derive the following set of equations
\begin{equation}\label{34}
    G^{\nu}_{\mu} + \Lambda g^{\nu}_{\mu} = \Sigma T^{\nu}_{\mu}
\end{equation}
\begin{equation}\label{35}
    -\frac{d-2}{2r} f'(r) - \frac{(d-2)(d-3)}{2r^2}(f(r)-1) - \Lambda = \Sigma T_{t}^{t} = \Sigma T_{r}^{r}
\end{equation}
\begin{equation}\label{36}
    -\frac{f''(r)}{2} - \frac{(d-3)}{2r} f'(r) - \frac{(d-3)(d-4)}{2r^2}(f(r)-1) - \Lambda = \Sigma T_{\theta_1}^{\theta_1} = \Sigma T_{\theta_{d-2}}^{\theta_{d-2}}
\end{equation}
As a consequence, the subsequent master equation is obtained
\begin{equation}\label{37}
    r^2 f''(r) + F_1 rf'(r) + F_2(f(r) - 1) + F_3 r^2 + F_4 r^{-2(d-3)} + F_5 r^{-(d-4)} = 0
\end{equation}
with
\begin{equation}\label{38}
\begin{split}
F_1 = (d - 1)\omega_q + 2(d - 5),\\
F_2 = (d - 3)((d - 1)\omega_q + d - 3),\\
F_3 = \frac{\Lambda}{2(d - 1)}(\omega_q + 1)^{\frac{d - 2}{2}},
\\
F_4 = q^2(d - 3)((d - 1)\omega_q - d + 3),
\\
F_5 = \frac{2((d - 1)\omega_q + 1)a}{d - 2}.
\end{split}
\end{equation}
It is important to note that the required advertising space cosmological constant $\Lambda$ is negative. Therefore, we can use Maxwell's equations $\nabla_\nu(\sqrt{-g}F^{\mu\nu}) = 0$ to evaluate
\begin{equation}\label{39}
    A = -\sqrt{\frac{(d-2)}{2(d-3)}} \frac{q}{r^{d-3}} dt.
\end{equation}
The solution to the primary equation is provided as follows
\begin{equation}\label{40}
    f(r) = 1-{\frac {m}{{r}^{d-3}}}+{\frac {{q}^{2}}{{r}^{2\,d-6}}}-2\,{\frac {
\lambda\,{r}^{2}}{ \left( d-2 \right)  \left( d-1 \right) }}-{\frac {
\alpha}{{r}^{ \left( d-1 \right) \omega+d-3}}}-2\,{\frac {a}{ \left( d
-2 \right) {r}^{d-4}}}
\end{equation}
here, "m" represents an integral constant that is directly proportional to the mass, while "q" represents a proportional constant related to the integral constant of BHs,The equation provided expresses the relationship described above \cite{r7,r8}.
\begin{equation}\label{41}
    M = \frac{(d-2)}{16\pi} \Omega_{d-2} m
\end{equation}
\begin{equation}\label{42}
    Q =\sqrt{\frac{2(d-2)(d-3)} \Omega_{d-2} q}{8\pi}
\end{equation}
here, the volume of a unit $(d-2)$-sphere, denoted as $\Omega_{d-2}$, and a +ve normalization factor called $\alpha$, which is associated with quintessence, play a crucial role. The relationship between $\alpha$ and the density $\rho_q$ can be found in \cite{r9}.
\begin{equation}\label{43a}
    \rho_q = -\frac{\alpha\omega_q(d-1)(d-2)}{4r^{d-1}(\omega_q+1)}
\end{equation}
Furthermore, the presence of the power $\left[ \frac{\alpha}{r^{d-1}\omega_q+d-3} \right]$ in Eq.(\ref{40}) can lead to distinct asymptotic effects of the quintessence term. In the case where only the quintessential contribution is taken into account, the aforementioned formula can be altered as follows
\begin{equation}\label{44a}
    f_{\alpha}(r) = 1 - \frac{m}{r^{d-3}} - \frac{\alpha}{r^{d-1}\omega_q+d-3}
\end{equation}
here, the spacetime exhibits an asymptotically de Sitter (dS)-like behavior when $\omega_q < -\frac{d-3}{d-1}$, while it becomes asymptotically flat otherwise and choose a specific value for $\omega_q$, namely $\omega_q = -\frac{d-2}{d-1}$, for numerical analysis.\\The mass can be defined as
\begin{equation}\label{51}
    M={\frac {\Omega_{{d-2}} \left( \left( d-1 \right) 
 \left( d-2 \right)(1/16\,{q}^{2} {r}^{-d+3}+1/16\,  {r}^{d-3}-1/16\,\alpha\, {r}^{d-2})+{r}^{d-1}\pi \,P-1/8\,ra \left( d-1 \right) 
 \right) }{\pi \, \left( d-1 \right) }}
\end{equation}
where, the mass of BH, represented by M, is defined to be its enthalpy. As a result, the equation provided expresses the relationship between enthalpy, internal energy, and pressure.
\begin{equation}
    M = U + P V.
\end{equation}
The BH`s entropy defined as
\begin{equation}
    S = \frac{1}{4} \Omega_{d-2} r^{d-2}
\end{equation}
\subsection{Topology}
\begin{figure}
    \centering
\includegraphics[width=8cm]{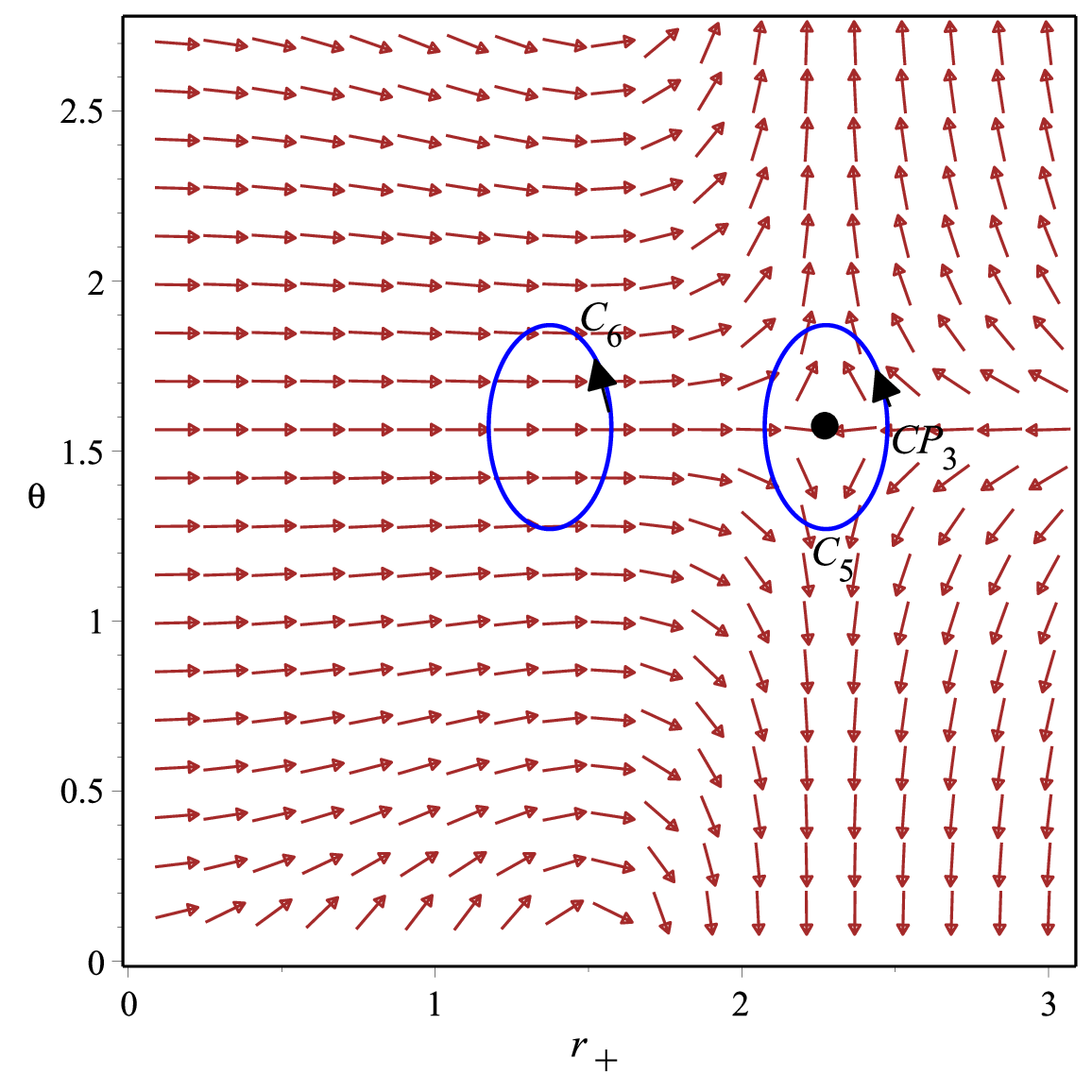}
    \caption{The brown arrows represent the vector field n of the $r_+-\theta$ plane for a D-dimensional RN-AdS BH with a cloud of string surrounding Quintessence.}
    \label{fig11}
\end{figure}

\begin{figure}
    \centering
\includegraphics[width=8cm]{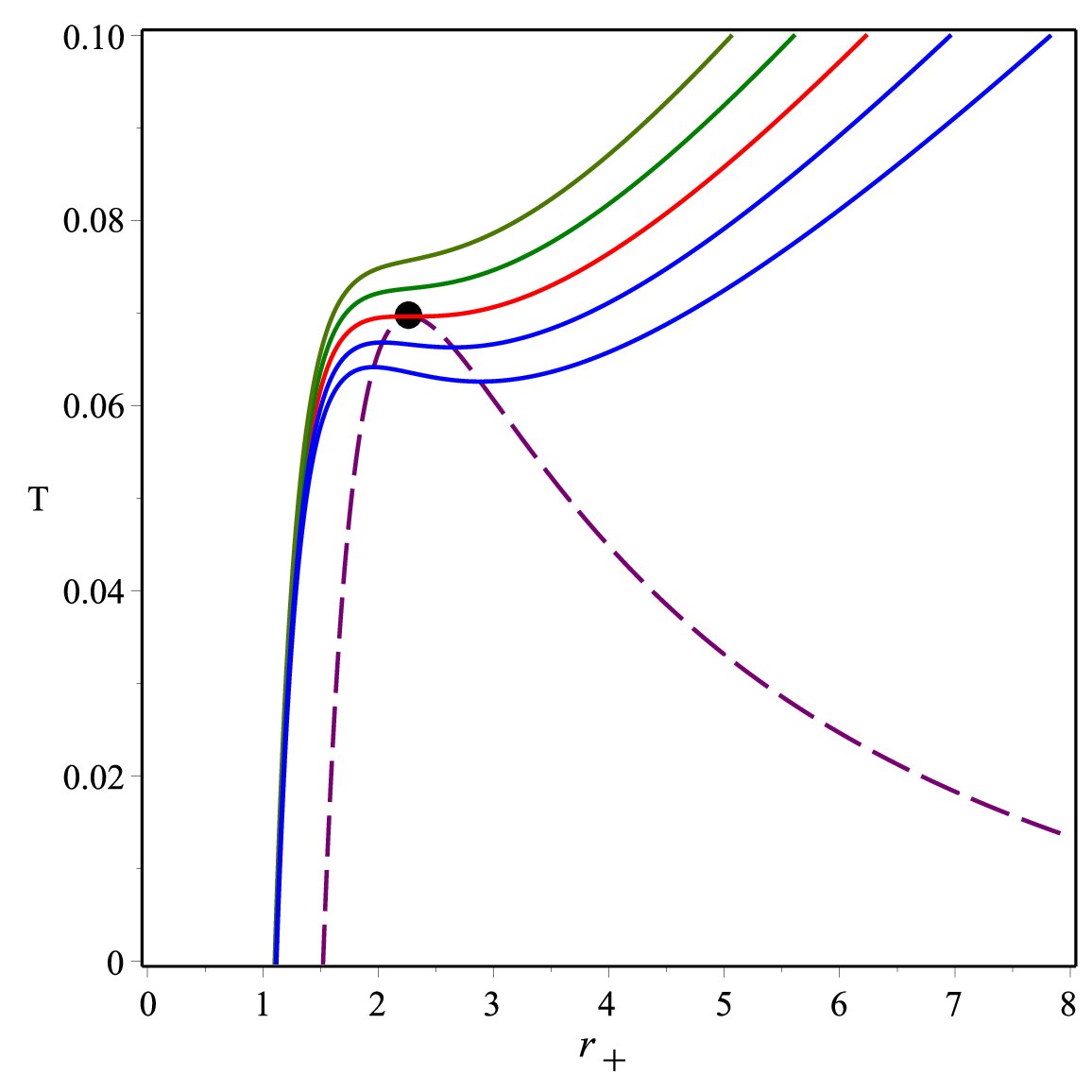}
    \caption{Isobaric curves (blue, green, and red) for a D-dimensional RN-AdS BH with a cloud of string surrounding Quintessence shown in $T-r_+$ plane. The black dot represents the critical point, and the purple dashed curve is for the extremal points of the temperature.}
    \label{fig12}
\end{figure}

To explore the thermodynamic topology of a D-dimensional RN-AdS BH with a surrounding cloud of string and Quintessence, it is possible to exhibits the temperature in relation to the charge, horizon radius $\&$ pressure.
\begin{equation}\label{13c}
    T = -{\frac {{r}^{-2\,d+6}{d}^{2}{q}^{2}-5\,{r}^{-2\,d+6}d{q}^{2}-16
\,\pi \,P{r}^{2}+\alpha\,{d}^{2}r+6\,{r}^{-2\,d+6}{q}^{2}-4\,\alpha\,d
r+2\,{r}^{-d+4}a+4\,\alpha\,r-{d}^{2}+5\,d-6}{4 \left( d-2 \right) r
\pi }}
\end{equation}
The critical points that corresponds are obtained as 
\begin{equation}\label{14c}
    \frac{\partial T}{\partial r_+} = 0, \frac{\partial^2T}{\partial^2r_+} = 0
\end{equation}
Use of the condition $(\frac{\partial_{r_+} T}{\partial_{r_+} S})_{q, P} = 0$, this results in a formula for the pressure, P.
\begin{equation}\label{15c}
    P=-{\frac { \left( d-3 \right)  \left(  \left( d-2 \right) {q}^{2
} \left( d-5/2 \right) {r}^{-2\,d+6}+{r}^{-d+4}a-d/2+1 \right) }{8{r}^{
2}\pi }}
\end{equation}
substituting P in Eq.(\ref{13c}), the term temperature, results in:
\begin{equation}\label{16c}
   T =  -{\frac {{q}^{2} \left( d-2 \right)  \left( d-3 \right) {r}^{-2\,
d+6}+1/2\,\alpha\,dr+{r}^{-d+4}a-\alpha\,r-d+3}{2 r\pi }}
\end{equation}
An updated thermodynamic function $\Phi$ is determined as,
\begin{equation}\label{17c}
    \Phi =  {\frac{1}{\sin{\theta}}}\,T(q,r_+)= -{\frac {{q}^{2} \left( d-2 \right)  \left( d-3 \right) {r}^{-2\,
d+6}+1/2\,\alpha\,dr+{r}^{-d+4}a-\alpha\,r-d+3}{2 \sin{\theta} r\pi }}
\end{equation}
The vector field`s $ \phi = (\phi^{r_+},\phi^\theta)$ components are
\begin{equation}\label{18c}
  \phi^{r_+} ={\frac { \left( {q}^{2} \left( d-2 \right)  \left( d-3 \right) {r
}^{-2\,d+6}+1/2\,\alpha\,dr+{r}^{-d+4}a-\alpha\,r-d+3 \right) \cos
 \left( \theta \right) }{ 2\left( \sin \left( \theta \right)  \right) ^
{2}r\pi }}
\end{equation}
and 
\begin{equation}\label{19c}
    \phi^{\theta} = {\frac { \left( d-3 \right)  \left( 2\, \left( d-2 \right) {q}^{2
} \left( d-5/2 \right) {r}^{-2\,d+6}+{r}^{-d+4}a-1 \right) }{2 \sin
 \left( \theta \right) {r}^{2}\pi }}
\end{equation}
The normalized vector $n = (\frac{\phi^{r_+}}{\|\phi\|},\frac{\phi^{\theta}}{\|\phi\|})$ is expressed in Fig.(\ref{fig11}). The plot exhibits the vector plot of $n$ in a $r_+$-$\theta$ plane for a D-dimensional RN-AdS BH with a cloud of string surrounding Quintessence. We have fixed $a = 0.1, \alpha = 1, d = 5, q = 1$ for this plot. The black dots represents the critical point $(CP_3)$, located at $(r_+,\theta) = (2.274,\frac{\pi}{2})$. \\
The topological charge associated with the critical point calculated as a contour $C$ is defined, parameterized by $\vartheta$ $\epsilon$ $(0,2\pi)$, as described in \cite{a8}.
\begin{equation}\label{22c}
     r_+ = \alpha\cos{\vartheta}+ r_o,\,
     \theta = \beta\sin{\vartheta}+ \frac{\pi}{2}
    \end{equation}
     Next, $C_5$ \& $C_6$, two contours, are created such that $C_5$ encloses $CP_3$, while $C_6$ does not. For such contours we take $(\alpha,\beta,r_o)$ = $(0.2,0.3,2.274)$ and $(0.2,0.3,1.184)$ respectively.\\
     The deviation experienced by the vector field $n$ as it follows the path along the contour $C$ can be characterized as:
   \begin{equation}\label{23c}
       \Omega(\vartheta) = \int^\vartheta_0 \epsilon_{\alpha\beta} n^\alpha \partial_ \vartheta n^\beta d\vartheta.
   \end{equation}
 The topological charge is defined as $Q = \frac{1}{2\pi} \Omega(2\pi)$. For $CP_3$, enclosed by $C_5$, the topological charge is determined to be $Q_{CP_3} = -1$, indicating a typical critical point. Since $C_6$ does not contain any critical point, so its topological charge is zero. Therefore, the overall topological charge is
   \begin{equation}
       Q = -1
   \end{equation}
 Notably, the critical radius is specified in the Eq.(\ref{14c}) precisely corresponds to the critical point identified through thermodynamic topology at coordinates $(r_+,\theta) = (2.274,\frac{\pi}{2})$, which has a topological charge of $-1$. In order to analyze the characteristics of the critical point, we present the surrounding phase structure (isobaric curves) in Figure (\ref{fig12}), where the critical point is indicated by a black dot. The red curve corresponds to the isobaric curve for $P = P_c$. Above the red curve, the green curves represent isobaric curves for $P > P_c$, while below the red curve, the blue curves represent isobaric curves for $P < P_c$. The purple dashed curve corresponds to the extremal points and is plotted using Eq. (\ref{16c}). By observing Figure.(\ref{fig12}), it can be noted that for $P < P_c$, the small $\&$ large BH phases are distinct by an unstable region, characterized by the negative slope of the isobaric curves or the area enclosed by the two extremal points associated with each isobaric curve. At the critical point, different phases of the D-dimensional RN-AdS BH with a surrounding cloud of string and Quintessence vanish. Therefore, the $CP_3$ can be regarded as a phase annihilation point, where one of the phases ceases to exist completely.
\subsection{Thermodynamic Topological Defects}
The study progresses to look at the D-dimensional RN-AdS BH solution with topological thermodynamic defects in the form of a cloud of string encircling Quintessence \cite{r10}. The BH's mass and entropy are used to derive an expression for the generalized free energy:
 \begin{equation}\label{24c}
     F = {\frac {1/16\,\left( d-1 \right)( \left( {q}^{2}t\left( d-2 \right) 
{r}^{-d+3}-\left( \alpha\, \left( d-2
 \right) t+4\,\pi  \right) {r}^{d-2}+ \left( \left( d-2 \right) {r}^{d-3}-2ra\right) t) +{r}^{d-1}\pi \,P\right) \Omega_{{d-2}}}{\pi \, \left( d
-1 \right) t}}
 \end{equation}
 The components of the vector field, as given by Eq. (\ref{7}), are:
 \begin{equation}\label{25c}
   \phi^{r_+} = \frac{1}{r \pi t} \left(\frac{\Omega_{d-2}}{16} q t (d-2)(d-3)r^{-d+3}\right) + \frac{1}{16} (d-2) (d-2) (\alpha (d-2) t + 4 \pi) r^{d-2} + \frac{1}{8} r^{d-3} \pi P - \frac{r}{8} \alpha t
\end{equation}
and
\begin{equation}\label{26c}
    \phi^\Theta = -\cot\Theta \csc\Theta
\end{equation}
\begin{figure}
    \centering
\includegraphics[width=8cm]{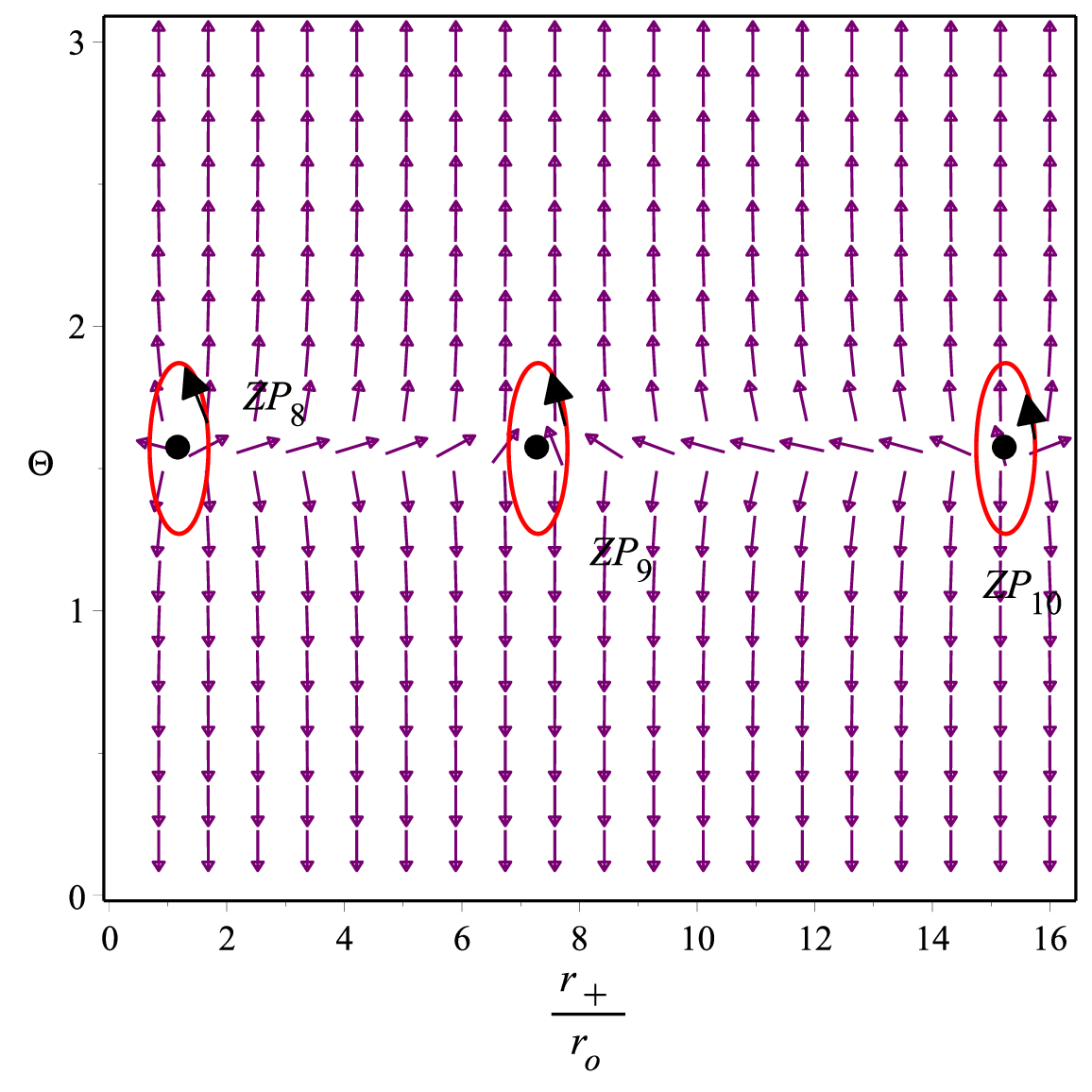}
    \caption{The purple arrows represent the vector field n of the $r_+/r_o-\Theta$ plane for $t/r_o = 150$ and P${r^2_o} = 0.001$ (P$<$ $P_c$). The zero point is indicated by the black dot.}
    \label{fig13}
\end{figure}
\begin{figure}
    \centering
\includegraphics[width=8cm]{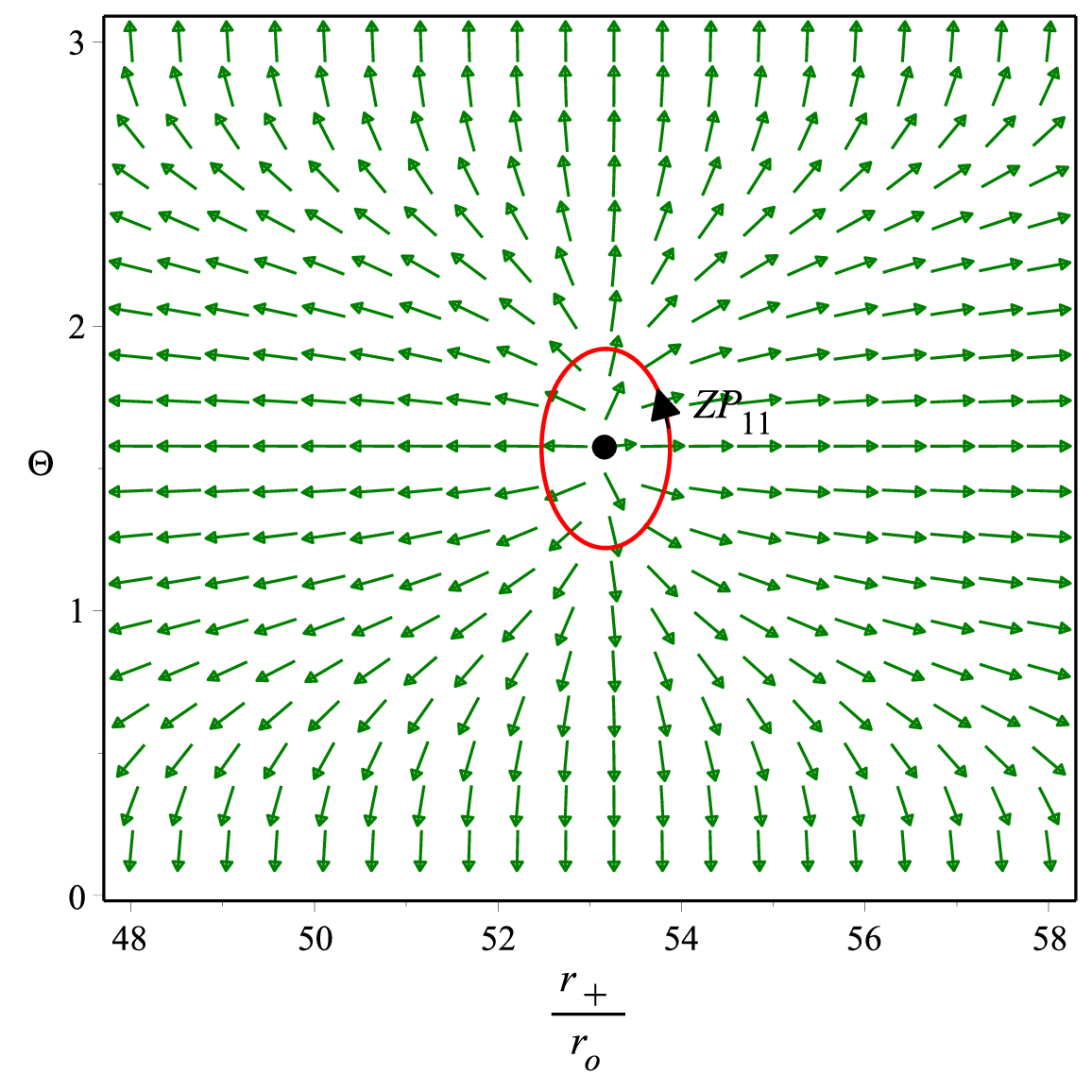}
    \caption{The purple arrows represent the vector field n of the $r_+/r_o-\Theta$ plane for $t/r_o = 20$ and P${r^2_o} = 0.001$ (P$<$ $P_c$). The zero point is indicated by the black dot.}
    \label{fig14}
\end{figure}
To determine the zero points, we plot and identify the unit vectors by putting $\Theta = \pi/2$ in $n^1$ for $a = 0.1, \alpha = 1, d = 5, q = 1$
and equating it equal to zero. For example, when $q/r_o = 1$, $t/r_o = 20$, and $Pr^2_o = 0.001$, we locate a zero point $(ZP_{11})$ at $(r_+/r_o, \Theta) = (53.17, \pi/2)$. The length scale $r_o$ is a non-specific value determined by the dimensions of the cavity that encloses the BH, while ensuring that the pressure remains below the critical pressure $P_c$.
 Figure.(\ref{fig14}) represents the unit vectors and the zero point. Similarly for $t/r_o = 150$, figure.(\ref{fig14}) represents the unit vectors and the zero points $(ZP_8)$,$(ZP_{9})$, and $(ZP_{10})$ with winding number $+1$, $-1$, and $+1$ respectively.Thus, the zero point's total topological charge or winding number is
\begin{equation}
    w=+1-1+1 = +1.
\end{equation}\\\\
An analytical expression for $t$ can be derived for the zero points by equating $\phi^{r_+}$ to zero.
\begin{equation}\label{29c}
    t = {\frac {4\pi \,{r}^{d-2} \left( d-2 \right) }{-{q}^{2} \left( d-2
 \right)  \left( d-3 \right) {r}^{-d+3}-\alpha\, \left( d-2 \right) ^{
2}{r}^{d-2}+ \left( {d}^{2}-5\,d+6 \right) {r}^{d-3}+16\,{r}^{d-1}\pi 
\,P-2\,ra}}
    \end{equation}\\
Figure. (\ref{fig15}) exhibits a graph of $r_+$ plotted against $t$, which was previously obtained. The curve's points represent the zeros of $\phi^{r_+}=0$. To generate this plot, we set $q/r_o = 1$, and $Pr^2_o = 0.001$ (P$<$ $P_c$).\\\\
\begin{figure}
    \centering
\includegraphics[width=8cm]{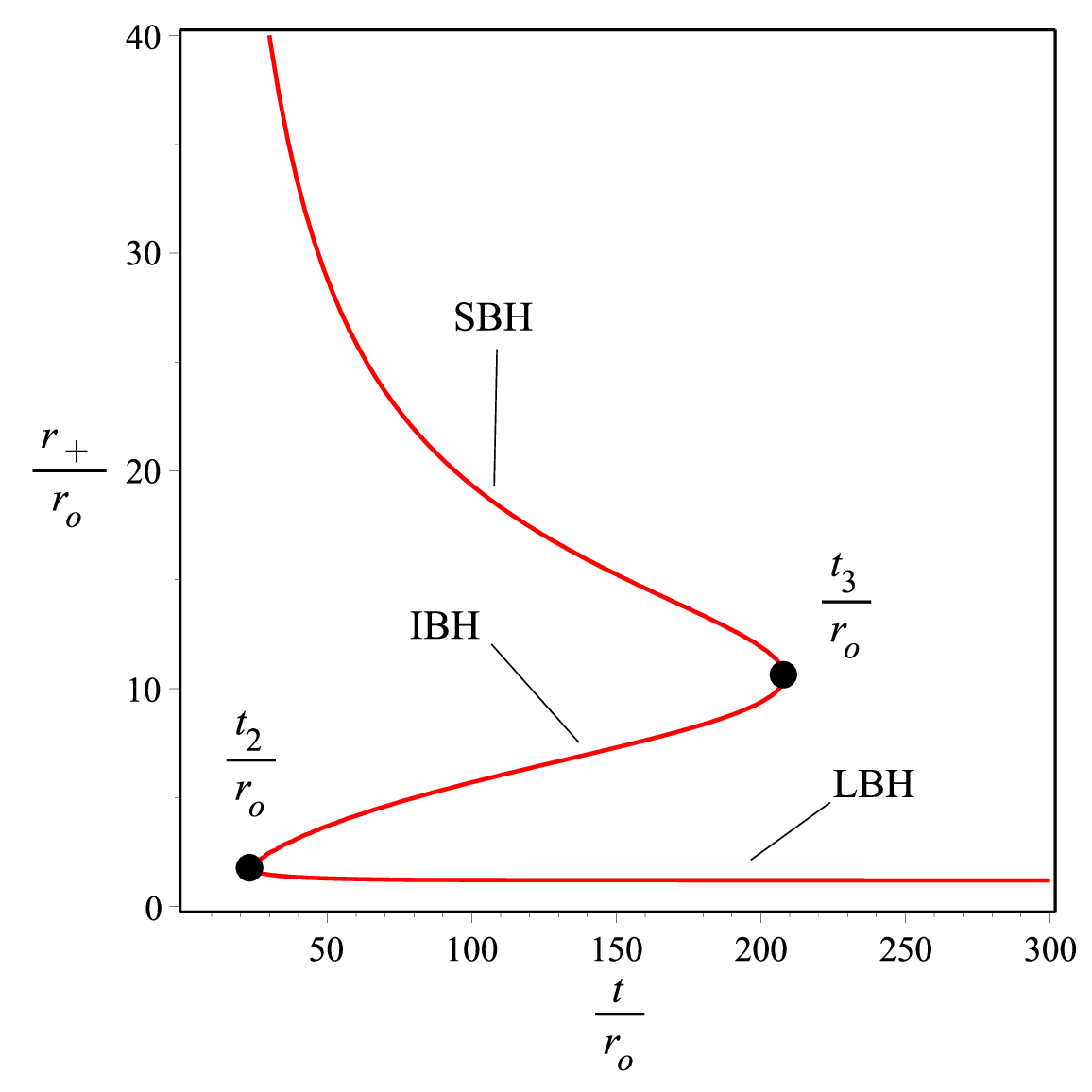}
    \caption{ The zero points of $\phi^{r_+}$ in $t/r_o$ vs. $r_+/r_o$ plane for D-dimensional RN-AdS BH with a cloud of string surrounding Quintessence}.
    \label{fig15}
\end{figure}
In Figure \ref{fig15}, three distinct branches of BHs can be clearly observed. The branch denoted by $t>t_3$ corresponds to the region of large BHs. It is notable that The determination of the winding number for any zero point along this branch yields a value of $w = +1$. Similarly, the branch indicated by $t<t_2$ corresponds to the region of small BHs, where any zero point along this branch exhibits an observed winding number of $w = +1$". The intermediate region of BHs is represented by the branch $t_2<t<t_3$, where the winding number for any zero point on this branch is found to be $w = -1$.
Therefore, the resulting topological number is $W = +1$. We performed explicit calculations of the specific heats for the three branches and discovered that the branches with a winding number of $+1$ exhibit positive specific heat, indicating thermodynamic stability. Conversely, the branch with a winding number of $-1$ displays negative specific heat, signifying thermodynamic instability.\\\\
Finally, the generation/annihilation points are determined by applying the given condition $\partial_{r_+}F =\partial_{r_+,r_+}F = 0$. For $q/r_o = 1$,  and $Pr^2_o = 0.001$. We find out  the generation and annihilation points at $t/r_o$ =  $t_2/r_o = 23.55 $ and $t/r_o$ = $t_3/r_o = 208.27$ respectively which are expressed by black dots in fig.\ref{fig15}.\\\\
Furthermore, we determine the zero points, plot and identify the unit vectors by putting $\Theta = \pi/2$ in $n^1$ for $a = 1, \alpha = 1, d = 5, q = 0.8$ and equating it to zero.For $t/r_o = 20$ we locate the zero point $(ZP_{12})$, with winding number $+1$. Figure. (\ref{fig16}) represents the unit vectors and the zero points for D-dimensional RN-AdS BH with a cloud of string surrounding Quintessence. Therefore, the zero point's total topological charge or  number is
\begin{equation}
    w=+1.
\end{equation}\\
Figure. (\ref{fig16}) right exhibits a graph of $r_+$ plotted against $t$, which was previously obtained. The curve's points represent the zeros of $\phi^{r_+}=0$. To generate this plot, we set $q/r_o = 1$, $a = 1, \alpha = 1, d = 5, q = 0.8$, and $Pr^2_o = 0.001$ (P$<$ $P_c$). The plot depicts that there is only one thermodynamically stable BH irrespect to the value of $t$.\\\\\\ 
\begin{figure}
    \centering
\includegraphics[width=8cm]{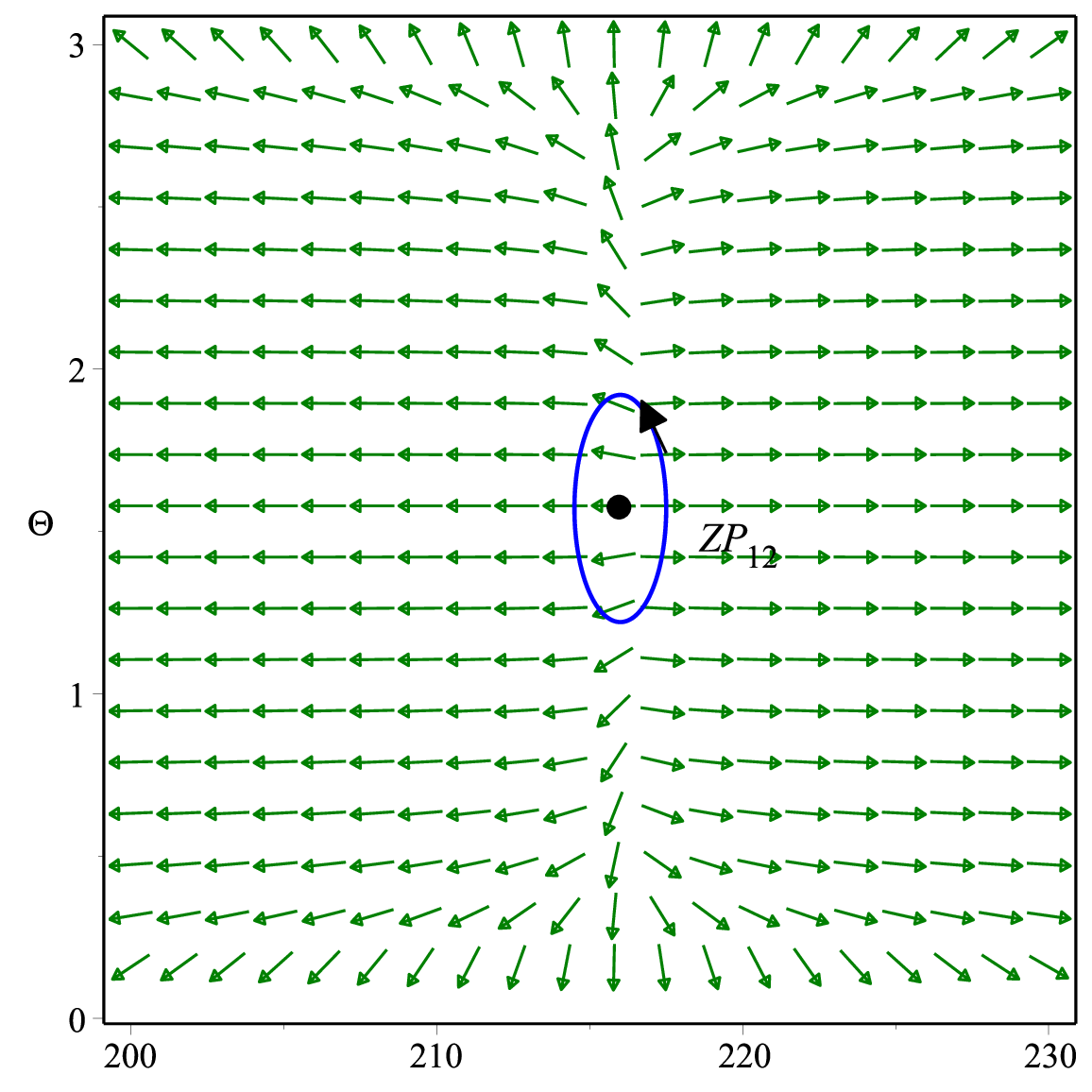}
\includegraphics[width=8cm]{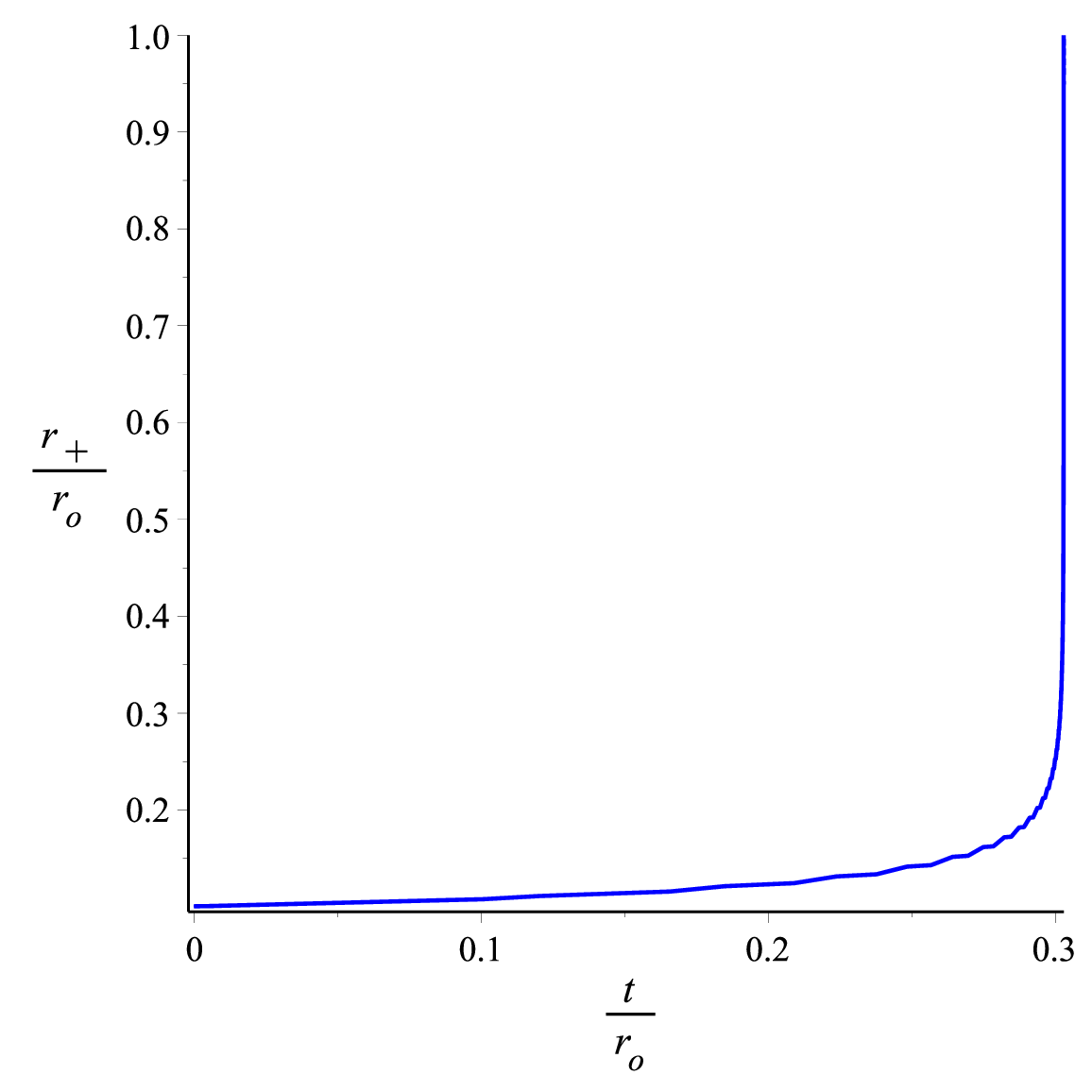}
    \caption{ Left: The vector field n of the $r_+/r_o-\Theta$ plane for $t/r_o = 20$ and P${r^2_o} = 0.001$ (P$<$ $P_c$). The zero point is indicated by the black dot. Right: The zero points of $\phi^{r_+}$ in $t/r_o$ vs. $r_+/r_o$ plane, at $a = 1, \alpha = 1, d = 5, q = 0.8$ for D-dimensional RN-AdS BH with a cloud of string surrounding Quintessence. }
    \label{fig16}
\end{figure}
\begin{table}[thb]
\begin{center}
 \renewcommand{\arraystretch}{2} 
 \newcommand*{\TitleParbox}[1]{\parbox[c]{1.75cm}{\raggedright #1}}%
\begin{tabular}{|c|c|c|c|c|}
\cline{1-2}
 \hline
 $\textbf{Black holes}$ & $\textbf{Cases}$ & $\textbf{Zero points}$ & $\textbf{Winding numbers}$ & \begin{tabular}{@{}ll@{}} \textbf{Total Topological}\\ \textbf{number} \end{tabular} \\
 \hline
 In dRGT massive gravity  &\begin{tabular}{@{}ll@{}} $\alpha = 2$, $\beta = 10$, $c = 1$ \\{$\alpha = -15$, $\beta = 10$, $c = 10$}
 \end{tabular} & \begin{tabular}{@{}ll@{}} $ZP_1$ \\ $ZP_2, ZP_3$ \end{tabular} &\begin{tabular}{@{}ll@{}} 1  \\ -1, +1 \end{tabular} &\begin{tabular}{@{}ll@{}} 1  \\ 0 \end{tabular} \\
 \hline
 In 5D Yang-Mills massive gravity & \begin{tabular}{@{}ll@{}}
 $c_1 = c_2 = c_3 = 1$, $e = 2$ \\  {$c_1 = 0.1, c_2 = -0.01, c_3 = 1$, $e = 0.2$} \end{tabular} &\begin{tabular}{@{}ll@{}} $ZP_4$\\$ZP_5$, $ZP_6$, $ZP_7$  \end{tabular}&
 \begin{tabular}{@{}ll@{}}  1\\ -1, +1, -1\end{tabular} &\begin{tabular}{@{}ll@{}}  1\\ 1\end{tabular} \\
 \hline
\begin{tabular}{@{}ll@{}}
 Quintessence surrounding \\  D-dimensional RN-Ads BH\\  with a cloud of string
 \end{tabular} 
 & \begin{tabular}{@{}ll@{}} $a = 0.1, \alpha = 1, d = 5, q = 1$\\\\$a = 1, \alpha = 1, d = 5, q = 0.8$ \end{tabular}
 & \begin{tabular}{@{}ll@{}} 
$ZP_8$, $ZP_9$, $ZP_{10}$  \\ $ZP_{11}$ &\\ $ZP_{12}$ \end{tabular}  &  \begin{tabular}{@{}ll@{}} +1, -1, +1\\+1\\+1
\end{tabular} & \begin{tabular}{@{}ll@{}} 1\\\\1 \end{tabular}
\\
 
 \hline
\end{tabular}
 \caption{Summary of the results}
\end{center}
\end{table}
\section{Conclusion}
In this work, we study the topological properties of BH in dRGT massive gravity, BH in five-dimensional Yang-Mills massive gravity and Quintessence surrounding a D-dimensional RN-AdS BH with a cloud of string. Our findings are summarized as
\begin{itemize}
    \item We find the critical points of these BHs, which are the points where the BHs undergo phase transitions between different phases. We also calculate their topological charges, which are quantities that measure the degree of non-triviality of the BH topology. We find that both BH in dRGT massive gravity and Quintessence surrounding a D-dimensional RN-AdS BH with a cloud of string have a single conventional critical point with a topological charge of -1. We plot the phase diagrams of these BHs in Figures.(\ref{fig1}) and (\ref{fig11}), where we mark the critical points by black dots.
    \item We treat the BHs as topological defects in the thermodynamic space, which is the space of thermodynamic variables such as temperature and pressure. We determine their topological numbers, which are sums of winding numbers of all defects. The winding number is an integer that indicates how many times a curve encircling a defect wraps around the origin. We find that the total topological number is equal to 1 for $\alpha = 2$, $\beta = 10$ and $c = 1$ and $0$ for $\alpha = -15$, $\beta = 10$ and $c = 10$ for the dRGT massive gravity model, 1 for all cases for the Yang-Mills massive gravity and RN-AdS BH with Quintessence and a cloud of strings model, respectively.
    \item We analyze the stability and the phase transitions of these BHs by using the specific heat and the free energy. We plot the specific heat curves in Figures. (\ref{fig2}) and (\ref{fig12}), where we mark the extremal points by black and purple dashed lines, respectively. The extremal points are the points where the specific heat diverges and indicate the boundaries between stable and unstable regions. We also examine from Figures.(\ref{fig7a}) and (\ref{fig15}) that there are three branches of BHs: small, intermediate, and large. The small and large BHs are stable for some parameter ranges, while the intermediate BHs are always unstable. The phase transitions between these BHs are similar to those of van der Waals liquid-gas system.
    \item We discuss the physical implications of our results. We show that the interchange of winding numbers can cause a phase transition by changing the topology of the order parameter space. This phenomenon has been observed in various systems such as superconductors, liquid crystals, magnets, etc., where it leads to changes in physical properties such as conductivity, magnetization, etc. This phenomenon can also lead to changes in the properties of BHs such as their mass, spin, etc., which can have significant implications for our understanding of these objects.
\end{itemize}

{\color{purple}

}
\end{document}